\documentclass[fleqn,usenatbib]{mnras}

\usepackage[T1]{fontenc}
\usepackage{ae,aecompl}
\usepackage{ragged2e} 
\usepackage{booktabs}
\usepackage{pdflscape} 
\usepackage[table,xcdraw]{xcolor}
\usepackage{multirow}
\usepackage{rotating,tabularx}

\usepackage{graphicx}    
\usepackage{amsmath}    
\usepackage{amssymb}    

\title[Chemical evolution of NGC 6809]{Chemical evolution of the metal poor Globular Cluster NGC 6809}

\author[M.J. Rain et al.]{
M.J. Rain,$^{1}$\thanks{E-mail: mariajoserain@gmail.com}
S. Villanova,$^{1}$
C. Mun\~oz$^{1}$
C. Valenzuela-Calderon$^{2}$
\\
$^{1}$ Departamento de Astronom\'ia, Universidad de Concepcion, Chile\\
$^{2}$ Departamento de Ingenier\'ia y Arquitectura, Universidad Arturo Prat, Chile\\
}

\date{Accepted XXX. Received YYY; in original form ZZZ}

\pubyear{2017}

\begin{document}
\label{firstpage}
\pagerange{\pageref{firstpage}--\pageref{lastpage}}
\maketitle

\begin{abstract}

We present the abundances analysis for a sample of 11 red giant branch stars in the metal-poor globular cluster NGC 6809 based on high-resolution spectra. Our main goals are to characterize its chemical composition and analyze this cluster's behavior associated with the Multiple Population (MPs) phenomenon. In our work we obtained the stellar parameters and chemical abundances of 24 elements (O, Na, Mg, Al, Si, Ca, Ti, V, Cr, Mn, Fe, Co, Sc, Ni, Cu, Zn, Y, Zr, Ba, La, Ce, Eu, Nd and Dy). We found a radial velocity of 174.7 $\pm$ 3.2 km $s^{-1}$ and a mean iron content of [Fe/H]=-2.01 $\pm$ 0.02 in good agreement with other studies. Moreover, we found a large spread in the abundances of the light elements O, Na and Al confirming the presence of a Na-O anti-correlation a Na-Al correlation. The Mg-Al anti-correlation is also present in our cluster. The $\alpha$ and iron-peak elements show good agreement with the halo field star trend. The heavy elements are dominated by the r-process.
\end{abstract}




\section{Introduction}
\label{Sec:intro}

Many photometric and spectroscopic evidence proved that GC host multiple populations (MPs). Star to star variations in light elements involved in the proton-capture process: C, N, O, Na, Mg and Al \citep{Osborn_1971, Cohen_1978, Cottrell_1981, Gratton_2001,  CarrettaGiraffe, CarrettaUves, Meszaros_2015,Schiavon_2017,Tang_2018} have been found among stars of the red-giant branch (RGB) or even on the main sequence (MS). It was suggested by \citet{Denisenkov_1989, Denisenkov_1990} that the origin of these variations occur during the Hydrogen burning at high temperature, involving several processes which are simultaneously active, where the processed material is transported to the surface of the star by means of the first dredge-up mixing process, occurring soon after the star leaves the MS. These processes are the CNO cycle ($\sim 15 \times 10^{6}$ K) which converts both C and O into N, the NeNa cycle (30$ \times 10^{6}$ K), where $^{20}$Ne is progressively converted in $^{23}$Na and the MgAl chain ($\sim 70 \times 10^{6}$ K) where $^{24}$Mg is finally converted in $^{26}$Al \citep{Denisenkov_1989,Prantzos_2007}.

One of the scenarios proposed to explain these patterns is the self- enrichment hypothesis. In this scenario, the patterns were inherited at the birth of the stars that we are currently observing. The discovery of the anti-correlations in the main sequence (MS) stars \citep{Gratton_2001} has given a new spin to the self-enrichment scenario. Since turnoff stars are not hot enough for the required nuclear reactions to occur in their interiors, the presence of light elements variations in these stars implies that the gas from which they formed has been polluted early in the history of the cluster by more massive evolving stars where O, Na, Mg and Al abundance anomalies pre-existed already. In this context, the spread present in light elements is believed to be a consequence of the presence of at least two generations of stars within the cluster that were formed during the first phase of the cluster evolution. Stars of the later generations formed from pristine gas polluted by hydrogen-burning processes material ejected by previous stars (the polluters) \citep{Prantzos_2006}. Several authors tried to describe the nature of these polluters. Among the most popular candidates there are massive asymptotic giant branch (AGB) stars  (M>4M$\odot$) \citep{Ventura_2001,Ventura_2016,Dellagli_2018}, fast rotating massive stars (FRMS), during their main sequence phase \citep{Decressin_2007}, massive binaries stars \citep{Demink_2009} and supermassive stars \citep{Denissenkov_2014}.

One of the most important traces of the existence of Multiple populations (MPs) is the anti-correlation between Sodium and Oxygen. This feature is present in all massive galactic GCs studied up to this date \citep[e.g][]{Yong_2008,Villanova_2010, Villanova_2016, Villanova_2017,Muñoz_2013, Muñoz_2017,Sanroman_2015, Mura_2017} and is considered as the footprint of the MPs \citep{Carretta_2010_Ca}. The only confirmed exception is Ruprecht 106 \citep{Villanova_Rup106}.\\

The detailed chemical analysis of globular cluster also provides information about the chemical evolution of the Galaxy and its star formation history. For example, a constant and enhanced [$\alpha$/Fe] indicate that the proto-cluster cloud from which the GC formed was not contaminated by thermonuclear supernovae (SNe). This implies that GCs formed within a couple of gigayears \citep{Carney_1996}. The abundance ratio of neutron capture elements (such as [Ba/Eu] and [La/Eu]) as a function of metallicity may also suggest how rapidly they were polluted by the low or intermediate mass stars before they were formed.

NGC 6809 also known as Messier 55 or M55 is a Halo GC in the constellation Sagittarius. It is located at 5.4 kpc ($l=8.79^{\circ}$, $b = -23.27^{\circ}$) from the Sun \citep{Harris}. The inferred mass for the cluster is 2.6$\times$10$^{5}$ \citep{Boyles}. NGC 6809 has an unusually low central concentration \citep[c=0.76,][]{Trager_1993} which makes it relatively easy to observe even within the cluster core. Due to its age, $\sim$ 13.5 $\pm$ 1.0 Gyr \citep{Dotter_2010}, $\sim$12.3 $\pm$ 1.7 Gyr \citep{Salaris_2002}, is a useful candidate to study the chemical evolution and star formation history of the Milky Way and despite the extensive photometric \citep{Richter_1999,Olech_1999,Kaluzny_2010,Rozy_2013}, kinematics and proper motion studies \citep[e.g][]{Zlo_2011,Sariya_2012} of this cluster, there is a lack of spectroscopic studies.\cite{Pilachowski_Alfa} \cite{ CarrettaGiraffe, CarrettaUves, Carretta_metallicity, Carretta_2010_Ca} and recently \cite{Wang} and \cite{Mucciarelli_2017} measured a number of elements ranging from Oxygen to Neodymium by using high-resolution spectra. However, none of these studies measured the large number of chemical elements that we were able to measure in our work. For this reason, in terms of the number of elements our study will be the most extensive and homogeneous in the past 20 years for NGC 6809.\\

\begin{figure}
    \includegraphics[width=1\columnwidth]{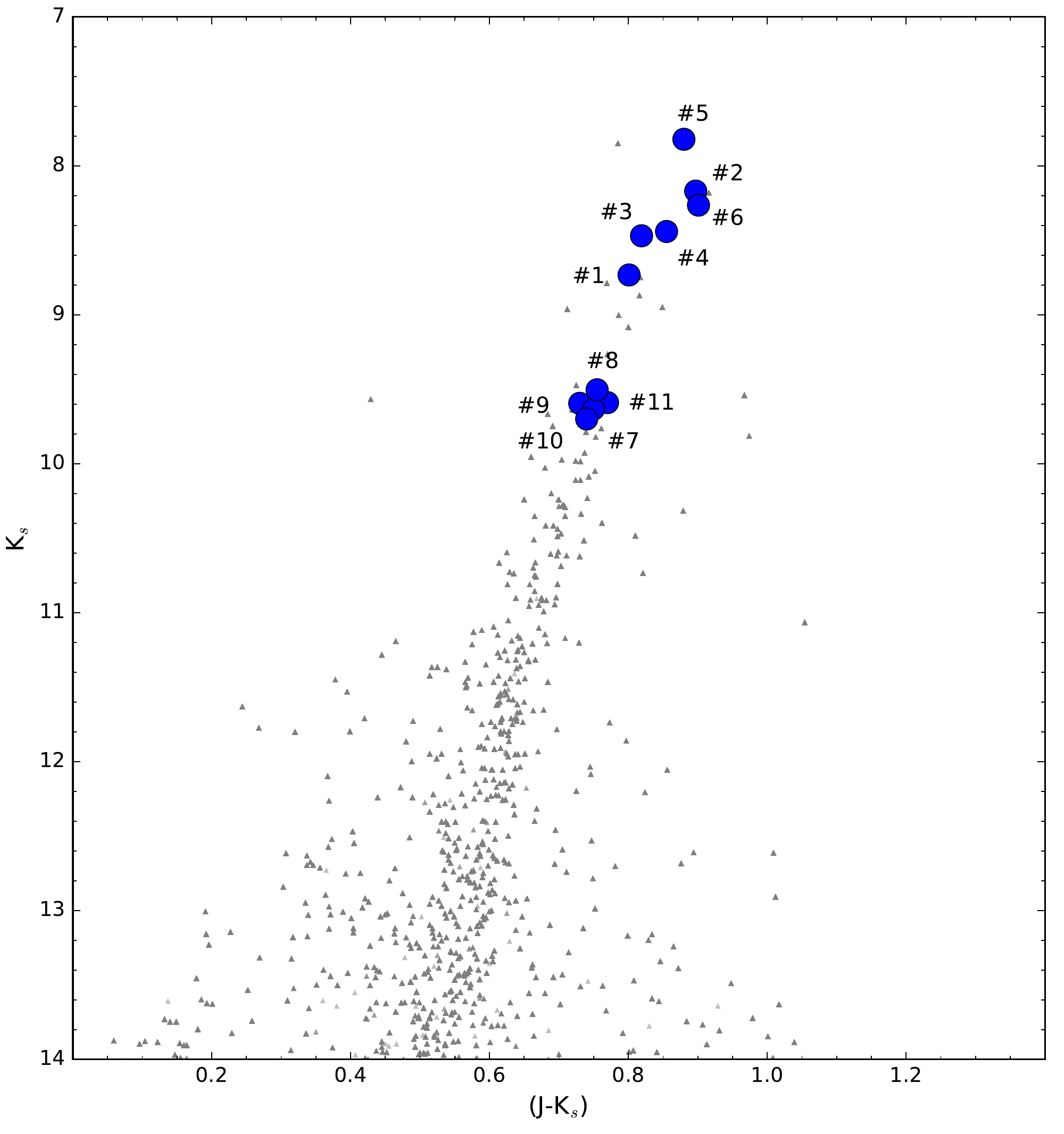}
    \caption{Color-Magnitude diagram (CMD) of NGC 6809 from 2MASS photometry. Blue circles are our observed UVES sample}
    \label{fig:cmd}
\end{figure}

This paper is organized in the following way:  A brief description about the data reduction is given in Section \ref{sec:obs_data}; In Section \ref{sec:abun_err}, we describe how to determine the stellar parameters, chemical abundances and errors. A comparison of our manual-derived chemical abundances and literature results are given in section \ref{sec:results}. Finally, a summary is given in Section \ref{sec:conclusions}.

\begin{table*}
    \centering
        \caption{Coordinates, 2MASS (J,H,K) magnitudes, heliocentric radial velocities, atmospheric parameters adopted and the typical S/N for the observed stars.}
\begin{tabular}{c c c c c c c c c c c c c}
 \hline 
 Star ID & RA & Dec & J$_{2MASS}$ & H$_{2MASS}$ & K$_{2MASS}$ & RV$_{H}$ & T$_{eff}$ & log(g) & v$_{t}$ & $[$Fe/H$]$ & S/N\\
 &[degrees]&[degrees]&[mag]&[mag]&[mag]&[Km$s^{-1}]$&[K]&[dex]&[Km$s^{-1}$]&[dex] &@6000 [\AA]\\
 \hline
 1 & 294.937416 & -31.003833 & 9.53 & 8.88 & 8.73 & 180.34 & 4209 & 0.14 & 2.02 &-2.21 & 165\\
 2 & 294.974874 & -30.960861 & 9.06 & 8.32 & 8.17 & 176.42 & 4105 & 0.28 & 1.86 &-1.93 & 136\\
 3 & 294.983416 & -30.921388 & 9.28 & 8.66 & 8.47 & 171.12 & 4144 & 0.06 & 2.13 &-1.97 & 159\\
 4 & 294.998541 & -30.919472 & 9.29 & 8.60 & 8.44 & 176.73 & 4162 & 0.51 & 1.76 &-1.94 & 142\\
 5 & 295.016999 & -30.972416 & 8.70 & 7.97 & 7.82 & 170.25 & 4022 & 0.08 & 2.07 &-1.93 & 114\\
 6 & 295.033583 & -30.971166 & 9.16 & 8.42 & 8.26 & 173.38 & 4119 & 0.30 & 1.80 &-1.93 & 123\\
 7 & 294.929999 & -30.947777 & 10.38 & 9.77 & 9.63 & 171.69 & 4376 & 0.66 & 1.81 &-2.05& 145\\
 8 & 294.960916 & -30.928388 & 10.25 & 9.61 & 9.50 & 175.85 & 4293 & 0.54 & 1.72 &-2.08& 72\\ 
 9 & 295.007583 & -30.979166 & 10.32 & 9.68 & 9.59 & 174.19 & 4350 & 0.61 & 1.77 &-2.04& 158 \\
 10 & 295.009999 & -30.785722 & 10.44 & 9.80 & 9.70 & 171.62 & 4387 & 0.63 & 1.75 &-2.01 & 60\\
 11 & 295.018124 & -31.062777 & 10.36 & 9.71 & 9.59 & 180.25 & 4348 & 0.65 & 1.77 &-2.06 & 200\\ 
  \hline  
  Cluster &  & & & & & 174.7 & & & & -2.01\\
  \hline
  Error &  & & & & & 3.26 & & & & 0.02\\
  \hline
      \end{tabular}
        \label{tab:basic_parameters}
\end{table*}

\section{Observations and Data reduction}
\label{sec:obs_data}

Our data-set  consists of high-resolution spectra collected at FLAMES/UVES spectrograph mounted on UT2 (Kueyen) at ESO-VLT Observatory in Cerro Paranal during June 2014 (ESO program ID 093.D-0286(A)). Our sample includes 11 stars, which belong to the Giant Branch cluster sequence. Infrared magnitudes were obtained from the Two Micron All-Sky Survey (2MASS) and range between $K_{s}$=7.97 and $K_{s}$=9.70. Figure \ref{fig:cmd} show the location of the members in the colour-magnitude diagram (CMD) J-K versus K.

The spectral coverage is $\sim$ 200nm (from 480 to 680nm), with the central wavelength at $\sim$ 580nm and with a mean resolution of R $\simeq$ 47000. We stacked several spectra in order to increment the signal-to-noise (S/N). The final S/N is between 60 and 200 at 600nm (see Table \ref{tab:basic_parameters}) 

Raw data were reduced using the dedicated pipeline\footnote{http://www.eso.org/sci/software/pipelines/}. Data reduction includes bias subtraction, flat-field correction, wavelength calibration and spectral rectification. We subtracted the sky using the \texttt{SARITH} package and radial velocities were measured by the \texttt{FXCOR} package in \texttt{IRAF} \footnote{IRAF is distributed by the National Optical Astronomy Observatory, which is operated by the Association of Universities for Research in Astronomy, Inc.., under cooperative agreement with the National Science Foundation}. This task cross-correlates the spectrum of the star with a template. Observed radial velocities were then corrected to the heliocentric system. NGC 6809 has a large heliocentric radial velocity RV$_{H}$. The mean value we obtained is $\langle RV_{H} \rangle$=174.7 $\pm$ 3.2 kms$^{-1}$, in excellent agreement with the literature. \cite{Harris} gives $\langle RV_{H} \rangle$ = 174.7 $\pm$ 0.3 kms$^{-1}$, \cite{Pryor_vrh} found a value of $\langle RV_{H} \rangle$ = 176 $\pm$ 0.9km$^{-1}$, \cite{Lane} provide a mean value of $\langle RV_{H} \rangle$ = 178 $\pm$ 1.21 km$^{-1}$ and finally \cite{Wang} found a radial velocity of $\langle RV_{H} \rangle$ = 173.6 $\pm$ 3.7 km$^{-1}$. We consider our value in agreement with the previous studies.\\

Table \ref{tab:basic_parameters} lists the basic parameters of the observed stars: the ID, J2000 coordinates (RA and Dec in degrees), J, H, K$_{s}$ magnitudes from 2MASS, heliocentric radial velocity RV$_{H}$ [kms$^{-1}$], adopted atmospheric parameters (including [Fe/H]) and the typical S/N for each star. In addition, we report the cluster mean radial velocity and mean [Fe/H] abundance with their errors. The determination of the atmospheric parameters is discussed in the following section.

\section{Atmospheric Parameters, Abundances and Error determination}
\label{sec:abun_err}
\subsection{Atmospheric parameters} \label{sub:model_at}
\begin{table*}
\centering
\caption{[X/Fe] for individual stars (columns 2 to 12) and the mean abundance ratios for the cluster (column 13). The abundances for Ti is the mean of those obtain from the neutral (TiI) and singly ionized species (TiII). The errors are statistical errors standard deviation of the mean. For Na abundances the NLTE values are reported.}
 \begin{tabular}{l c c c c c c c c c c c c c c c c c}
 \hline 
 [X/Fe] & 1 & 2 & 3 & 4  & 5 & 6 & 7 & 8 & 9 & 10 & 11 & $\langle[$X/Fe$]\rangle$  & Sun\\
  \hline 
 $[$O/Fe$]$   & +0.43 & +0.08 & +0.38  & +0.09 & +0.21 & +0.13 & -0.17 & +0.13 & -0.02 & -0.19 & +0.18 & +0.11$\pm$0.05 &  8.83\\
 $[$Na/Fe$]$  & -0.05 & +0.53 & -0.06  & +0.52 & +0.25 & +0.42 & +0.53 & +0.46 & +0.44 & +0.33 & +0.36 & +0.33$\pm$0.06 &  6.32\\
 $[$Mg/Fe$]$  & +0.48 & +0.51 & +0.55  & +0.24 & +0.31 & +0.45 & +0.27 & +0.53 & +0.41 & +0.39 & +0.45 & +0.41$\pm$0.03  &  7.56\\
 $[$Al/Fe$]$  & +0.32 & +0.95 & +0.16  & +1.01 & +0.82 & +1.00 & +1.24 & +0.79 & +1.18 & +1.32 & +0.72 & +0.86$\pm$0.10  &  6.43\\
 $[$Si/Fe$]$  & +0.50 & +0.50 & +0.59  & +0.39 & +0.48 & +0.53 & +0.49 & +0.54 & +0.59 & +0.45 & +0.48 & +0.50$\pm$0.01 &  7.61\\
 $[$Ca/Fe$]$  & +0.40 & +0.31 & +0.39  & +0.30 & +0.33 & +0.30 & +0.37 & +0.36 & +0.36 & +0.28 & +0.36 & +0.34$\pm$0.01 &  6.39\\
 $[$Sc/Fe$]$  & +0.19 & +0.18 & +0.14  & +0.13 & +0.21 & +0.21 & +0.08 & +0.08 & +0.10 & +0.06 & +0.12 & +0.13$\pm$0.01  & 3.12\\
 $[$Ti/Fe$]$  & +0.37 & +0.33 & +0.36  & +0.41 & +0.38 & +0.49 & +0.37 & +0.33 & +0.34 & +0.24 & +0.34 & +0.36$\pm$0.02  &  4.92\\
 $[$V/Fe$]$   & +0.08 & +0.10 & +0.09  & +0.10 & +0.14 & +0.14 & +0.15 & -0.02 & +0.06 & +0.19 & +0.16 & +0.10$\pm$0.02 &  4.00\\
 $[$Cr/Fe$]$  & -0.01 & -0.03 & -0.10  & -0.01 & +0.10 &  0.00 & -0.01 & +0.04 & +0.02 & -0.08 & +0.03 &  0.00$\pm$0.02  &  5.63\\
 $[$Mn/Fe$]$  & -0.29 & -0.25 & -0.28  & -0.24 & -0.21 &  0.24 & -0.39 & -0.30 & -0.33 &........&-0.29 & -0.28$\pm$0.01  & 5.37\\
 $[$Co/Fe$]$  & +0.32 & +0.12 & +0.20  & +0.10 & +0.12 & +0.10 & +0.19 & +0.19 & +0.17 &........&+0.24 & +0.17$\pm$0.02  & 4.93\\
 $[$Ni/Fe$]$  & +0.01 & -0.04 & -0.03  & -0.02 & -0.03 & -0.02 & +0.01 & +0.01 & +0.04 & -0.02  &+0.01 &  0.00$\pm$0.01 &  6.26\\
 $[$Cu/Fe$]$  & -0.13 & -0.28 & -0.29  & -0.32 & -0.40 & -0.37 & -0.19 & -0.27 &  0.22 &........&-0.34 & -0.28$\pm$0.02 & 4.19\\
 $[$Zn/Fe$]$  & +0.36 & +0.23 & +0.27  & +0.16 & +0.33 & +0.28 & +0.09 & +0.25 & +0.11 & +0.07  &+0.16 & +0.22$\pm$0.03  & 4.61\\
 $[$Y/Fe$]$   & +0.24 & +0.12 & -0.02  & +0.10 & +0.13 & +0.07 & -0.07 & -0.02 & -0.08 & -0.12  &-0.08 & +0.02$\pm$0.04  & 2.25\\
 $[$Zr/Fe$]$  & +0.39 & +0.13 & +0.24  & +0.31 & +0.26 & +0.19 & +0.17 & +0.25 & +0.15  &........&+0.22&+0.23$\pm$0.02  & 2.56\\
 $[$Ba/Fe$]$  & +0.03 & -0.09 & +0.13  & -0.03 & +0.05 & -0.07 & -0.18 & -0.19 & -0.21  & -0.15 &-0.21 &-0.08$\pm$0.03  & 2.34\\
 $[$La/Fe$]$  & +0.35 & +0.24 & +0.18  & +0.27 & +0.29 & +0.25 & +0.23 & +0.17 & +0.16  & +0.11 &+0.19 &+0.22$\pm$0.02  & 1.26\\
 $[$Ce/Fe$]$  & -0.05 & -0.08 & -0.17  & -0.11 & +0.07 & -0.07 & -0.06 & -0.11 & -0.18  &........&-0.14 &-0.09$\pm$0.02 & 1.53\\
 $[$Nd/Fe$]$  & +0.40 & +0.31 & +0.26  & +0.36 & +0.43 & +0.39 & +0.30 & +0.25 & +0.25  &  +0.22&+0.30 &+0.31$\pm$0.02  & 1.59\\
 $[$Eu/Fe$]$  & +0.64 & +0.54 & +0.47  & +0.63 & +0.59 & +0.63 & +0.42 & +0.54 & +0.57  & +0.41 &+0.56 &+0.54$\pm$0.02  & 0.52\\
 $[$Dy/Fe$]$  & +0.30 & +0.13 & +0.05  & +0.17 & +0.17 & +0.25 & +0.25 & +0.15 & +0.16  &........&-0.06&+0.15$\pm$0.03 & 1.10\\
\hline 
[$\alpha$/Fe]& +0.43 & +0.41 & +0.47 & +0.33 & +0.37 & +0.44 & +0.37 & +0.44 & +0.42 & +0.34 & +0.40 & +0.40$\pm$0.01 \\
\hline
\end{tabular}
\label{tab:abundances}
 \end{table*}

 A first estimation of stellar parameters was obtain by using infrared photometry from 2MASS and using the following procedure:  effective temperature was derived from the (J-K) color using the relation by \citet{Alonso} and the reddening E(B-V)=0.08 from \citet{Harris}. Surface gravity log(g) was obtained from the canonical equation:
 
\begin{equation} 
\centering
log\left(\frac{g}{g_{\odot}}\right) =log\left(\frac{M}{M_{\odot}}\right)+ 4 log\left(\frac{T_{eff}}{T_{\odot}}\right) - log\left(\frac{L}{L_{\odot}}\right)
\end{equation}

where the mass M was assumed to be 0.8M$_{\odot}$ and the luminosity L/L$_{\odot}$ was obtained from the absolute magnitude M$_{V}$ assuming a distance modulus of (m-M)$_{0}$= 13.89 from \citet{Harris}. The bolometric correction (BC) was derived by adopting the BC - T$_{eff}$ relation from \citet{Alonso}. Finally, microturbulence velocity (v$_{t}$) was obtained from the relation v$_{t}$ - log(g) from \citet{Gratton_vt} 

\begin{equation}
v_{t}=2.22-0.322 \cdot log(g)
\end{equation}
These atmospheric parameters are only initial guesses and were then adjusted in order to calculate the final atmospheric parameters using \texttt{MOOG}.

\texttt{MOOG} \citep{Sneden_1973} is a \texttt{FORTRAN} code that performs a spectral lines analysis assuming Local Thermodynamic Equilibrium (LTE). The line list for the chemical analysis is the one described in \cite{Line_villa}).

For this analysis, the model atmosphere was initially calculated using the \texttt{ATLAS9} code \citep{Kurucz} adopting the values of T$_{eff}$, v$_{t}$ obtained from photometry and using [Fe/H]=-1.94 dex value from \citet{Harris}.  Then, during the abundance analysis, these values were adjusted by imposing an excitation potential (EP) equilibrium to the FeI lines for T$_{eff}$. Thus minimising any dependence of FeI abundances from equivalent widths (EWs) we obtained the final v$_{t}$ and finally for log(g) was derived satisfying the ionisation equilibrium of species ionised differently (FeI and FeII lines). The [Fe/H] value of the model was changed at each iteration according to the output of the abundance analysis. The adopted atmospheric parameters are listed in Table \ref{tab:basic_parameters}.

\subsection{Chemical abundances}
\label{Sub:chemica}

\begin{figure}
    \includegraphics[width=1\columnwidth]{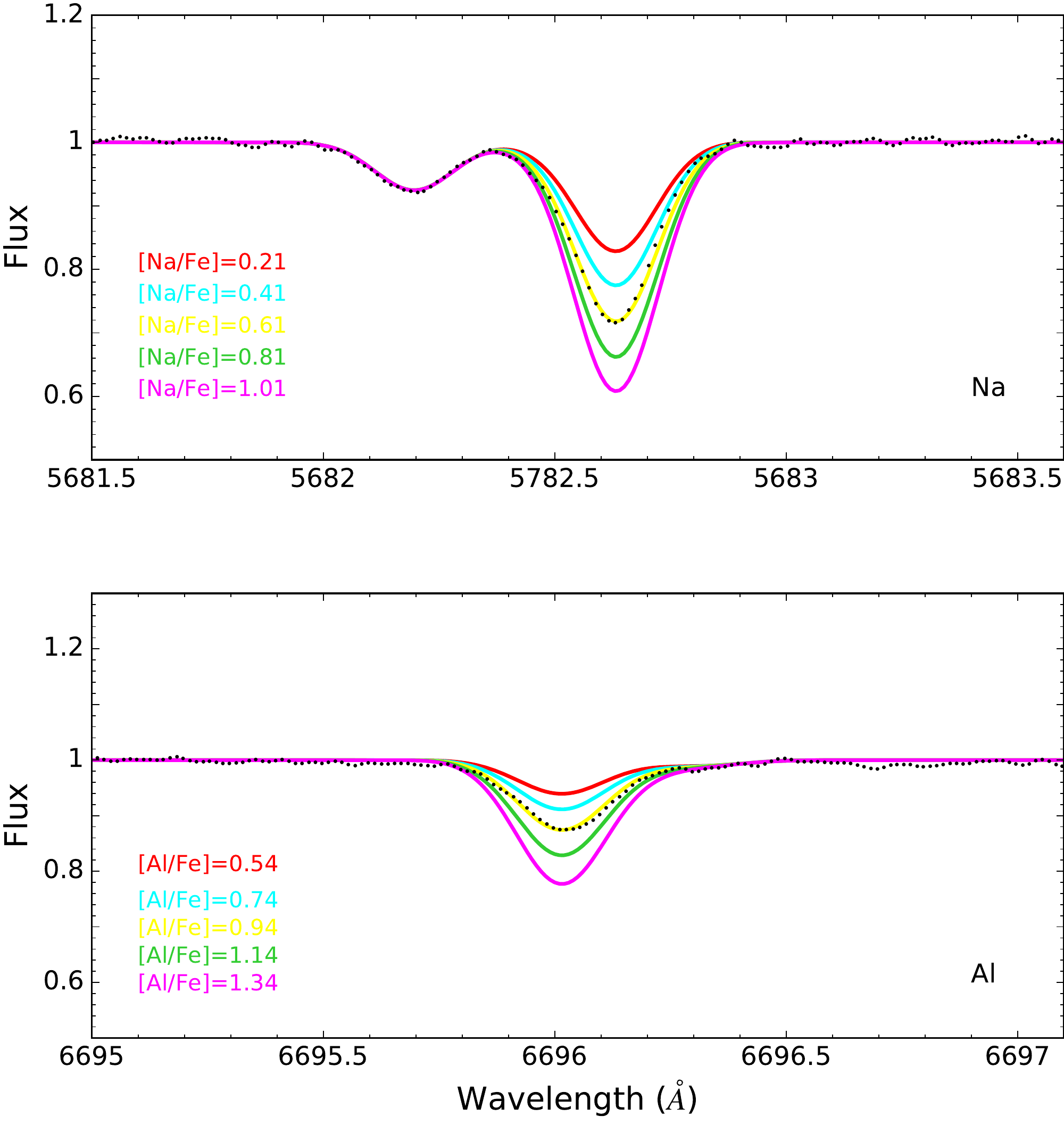}
    \caption{Sodium and Aluminum lines fitting process in star \#2. \textbf{Dotted} lines correspond to observed lines and \textbf{solid} color lines are synthesized spectra for different abundances.}
    \label{fig:MOOG}
\end{figure}

In our study, we used two different methods to measure abundances: equivalent widths (EW's) and spectral synthesis. Isolated well-defined lines were treated with the former. A detailed explanation of the method used to measure the EW's is given in \citet{Marino_2008}. For those elements whose lines are weak or affected by blending or hyperfine and isotopic splitting, we used the spectrum-synthesis method. Abundances for our targets were determined by using the standard running option \texttt{SYNTH} (one of the typical task in determination of chemical abundances in  \texttt{MOOG}). The chemical abundances in this paper where derived by visual inspection where the synthetic spectrum is compared to the observed one modifying the abundance at each step until reaching the best-fit value as the one that minimises the rms scatter. An example of a spectral synthesis is plotted in Figure \ref{fig:MOOG} where the spectrum of the observed star \#2 for Na and Al absorption lines is compared with synthetic spectra.

In this work, we measured the light elements O, Na, Mg, Al. Both Oxygen and Sodium abundances were obtained using the spectral synthesis method. For O we used the forbidden [OI] line at 6300 \AA, for Na abundances we used the Na I doublet at 5682 - 5688 \AA. Non-LTE (NLTE) corrections were applied to Na lines; this will be discussed in more detail in Subsection \ref{Sub:Na-O}. Magnesium abundances are typically derived from three high excitation lines: 5711, 6318 and 6319 \AA. However, in our study Mg I abundances have been inferred only from spectral synthesis only of the transitions at $\sim$ 5711.09 \AA. Aluminum abundances are derived from the doublet at 6696 - 6698 \AA. 

Iron abundances were estimated through the EW's method using 38-51 lines  (depending on the S/N) for FeI and 5 lines for FeII. For Calcium and Titanium 5 and 15 lines were used respectively.
 Silicon abundances were inferred from several transitions within the spectral range 5645-61445 \AA. SiI presents few features in the spectrum, so in this case, abundances derived from EW's were cross-checked with the spectral synthesis method in order to obtain more accurate measurements. 

By using spectral synthesis we determined abundances of the following iron peak elements: Sc (5684 \AA) ,CuI (5218 \AA), Co (5342 \AA), VI (5670 \AA), MnI (5420 \AA), and ZnI (4811 \AA). For Nickel and Chromium abundances we used a number of 17 - 23 and 10 isolated lines respectively. NiI and CrI abundances are all derived by EW's method. Manganese, Cobalt and Copper abundances are available only for ten stars.

Finally, to complete our analysis and include key heavy elements the abundances abundances, of Y, Zr, La, Ce, Nd, Eu, Ba and Dy were obtained. For this set of elements, only the spectrum-synthesis method was used. Even if several of these elements suffer hyperfine splitting only Ba lines, being strong, need this splitting to be taken into account. For Zr II we used the available lines at 5112 and 5350 \AA. Zirconium, Dysprosium and Cerium abundances are available only for ten stars. A list of all the derived chemical abundances and the adopted solar abundances we used are reported in Table \ref{tab:abundances}. 

\begin{figure*}
    \includegraphics[width=2.1\columnwidth]{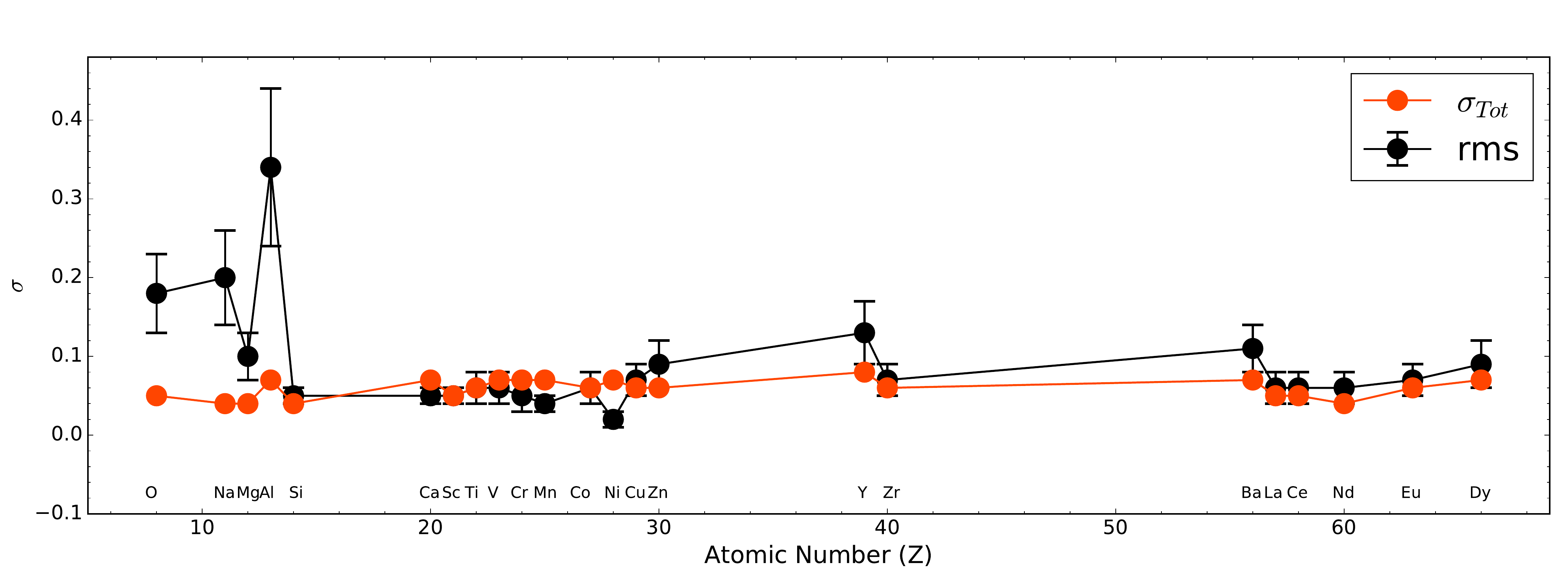}
    \caption{ rms scatter and $\sigma_{Tot}$ versus Atomic number (Z). Black filled circles represent the mean rms scatter and red filled circles represent the mean total error from our sample. Error bars represent the error of the mean.}
    \label{fig:errors}
\end{figure*}

\subsection{Errors}
\label{Sub:errors}

To estimate the internal errors associated with our spectroscopic atmospheric parameters we performed the following procedure.

For T$_{eff}$ we calculated, in the relation between abundance versus potential excitation (E.P), the errors associated with the slopes by performing the best least square fit to each star. The average of the errors corresponds to the typical error on the slope. Then we selected three stars representative of the entire sample star\#10, star\#4, and star\#5  with high, intermediate and low T$_{eff}$ respectively. For each of them, we fixed log(g), v$_{t}$ and [Fe/H] and varied the temperature until the slope of the line that best fits the relation between abundances and E.P. given by  \texttt{MOOG} became equal to the respective mean error we find. The difference in temperature represents an estimate of the error in temperature itself.
In order to associate an error to the measures of gravity, we have varied the gravity of the three representative stars such that the relation

$$[Fe/H]-\langle\sigma_{star} [FeI/H]\rangle= [Fe/H]+\langle\sigma_{star} [FeII/H]\rangle $$
was satisfied for each. Where $\langle\sigma_{star} [Fe/H]\rangle$ is the dispersion given by \texttt{MOOG} for the corresponding lines, divided by the square of the number of lines minus one \citep{Marino_2008}.
For v$_{t}$ the same procedure as in temperature was applied, but in this case, using the relation between abundance and EWs. 
The error of the mean [Fe/H] what obtained by dividing the rms ($\sigma_{obs}$) by $\sqrt{N_{lines}-1}$. 

We find $\Delta$T$_{eff}$=40k, $\Delta$v$_{t}$=0.07 kms$^{-1}$, $\Delta$log(g)= 0.10 and $\Delta$[Fe/H]= 0.02 dex as our errors on the atmospheric parameters. Then we choose star \#7 as representative of the sample, varied its T$_{eff}$, log(g), [Fe/H], and v$_{t}$ according the atmospheric errors we just obtained, and redetermining the abundances.

Finally, we measured the error due to the noise in the spectra for each star. This error was obtained for elements whose abundance was obtained by EWs, as the errors on the mean given by \texttt{MOOG}. On the other hand, elements whose abundances was obtained by spectral synthesis the error is the output of the fitting procedure. The final errors ($\sigma_{Tot}$) in our measured abundances are given by the relation:

\begin{equation}
\label{eq:error_final}
\sigma_{tot}=\sqrt{\sigma^{2}_{T_{eff}} +\sigma^{2}_{log(g)} +\sigma^{2}_{v_{t}}+ \sigma^{2} +\sigma^{2}_{[Fe/H]} + \sigma^{2}_{S/N}}
\end{equation}

 The resulting errors for each [X/Fe] ratio, including the observed dispersion ($\sigma_{Obs}$) and the error of the S/N are reported in Table \ref{tab:error}. In Figure \ref{fig:errors} we plot the rms scatter and total errors as a function of the atomic number for each chemical element. Error bars represent the error of the mean rms scatter (rms/$\sqrt{N}$, where N is the number of data used to obtain the rms). The agreement between $\sigma_{Tot}$ and the rms scatter is good for all elements except for O, Na, Mg and Al.

\begin{table*}

\centering
\caption{Estimated errors on [X/Fe] due to the stellar parameters (columns 2 to 5) and to spectral noise (column 6) for star \#7. Column 7 is the total internal uncertainty calculated as the root square of the sum of the square of columns 2 to 6. Column 8 is the rms scatter.}
 \begin{tabular}{l c c c c c c c}
 \hline 
[X/Fe] & $\Delta$ T$_{eff}$ & $\Delta$log(g)  & $\Delta v_{t}$  & $\Delta$[Fe/H] & $\sigma_{S/N}$ & $\sigma_{Tot}$ & $\sigma_{Tot}$ (rms)\\
&40k& 0.06 kms$^{-1}$  & 0.10 & 0.02 dex  & & \\
 \hline
$\Delta([$O/Fe$])$   & 0.03 &               0.04 &  0.02 &    0.01 &       0.01 &      0.05 &     0.18\\
$\Delta([$Na/Fe$])$  & 0.03 &               0.01 &  0.03 &    0.01 &       0.02 &      0.04 &     0.20\\
$\Delta([$Mg/Fe$])$  & 0.02 &               0.03 &  0.01 &    0.03 &       0.01 &      0.04 &     0.10\\
$\Delta([$Al/Fe$])$  & 0.06 &               0.03 &  0.04 &    0.00 &       0.01 &      0.07 &     0.34\\
$\Delta([$Si/Fe$])$  & 0.02 &               0.03 &  0.02 &    0.01 &       0.02 &      0.04 &     0.05\\
$\Delta([$Ca/Fe$])$  & 0.06 &               0.00 &  0.01 &    0.02 &       0.04 &      0.07 &     0.05\\
$\Delta([$Sc/Fe$])$  & 0.00 &               0.05 &  0.00 &    0.00 &       0.01 &      0.05 &     0.05\\
$\Delta([$TiI/Fe$])$ & 0.08 &               0.00 &  0.01 &    0.01 &       0.02 &      0.08 &     0.04\\
$\Delta([$TiII/Fe$])$& 0.01 &               0.04 &  0.01 &    0.00 &       0.04 &      0.05 &     0.09\\
$\Delta([$V/Fe$])$   & 0.07 &               0.01 &  0.02 &    0.04 &       0.02 &      0.07 &     0.06\\
$\Delta([$Cr/Fe$])$  & 0.06 &               0.00 &  0.01 &    0.01 &       0.05 &      0.07 &     0.05\\
$\Delta([$Mn/Fe$])$  & 0.02 &               0.05 &  0.05 &    0.04 &       0.01 &      0.07 &     0.04\\
$\Delta([$Fe/H$])$   & 0.03 &               0.03 &  0.04 &    0.03 &       0.01 &      0.07 &     0.08\\
$\Delta([$Co/Fe$])$  & 0.06 &               0.01 &  0.03 &    0.01 &       0.01 &      0.06 &     0.06\\
$\Delta([$Ni/Fe$])$  & 0.07 &               0.00 &  0.01 &    0.01 &       0.02 &      0.07 &     0.02\\
$\Delta([$Cu/Fe$])$  & 0.05 &               0.02 &  0.04 &    0.01 &       0.01 &      0.06 &     0.07\\
$\Delta([$Zn/Fe$])$  & 0.03 &               0.03 &  0.02 &    0.02 &       0.04 &      0.06 &     0.09\\
$\Delta([$Y/Fe$])$   & 0.05 &               0.05 &  0.04 &    0.02 &       0.03 &      0.08 &     0.13 \\
$\Delta([$Zr/Fe$])$  & 0.03 &               0.03 &  0.03 &    0.02 &       0.01 &      0.06 &     0.07\\
$\Delta([$Ba/Fe$])$  & 0.04 &               0.03 &  0.03 &    0.04 &       0.03 &      0.07 &     0.11\\
$\Delta([$La/Fe$])$  & 0.03 &               0.03 &  0.02 &    0.02 &       0.01 &      0.05 &     0.06\\
$\Delta([$Ce/Fe$])$  & 0.04 &               0.02 &  0.01 &    0.02 &       0.01 &      0.05 &     0.06\\
$\Delta([$Nd/Fe$])$  & 0.02 &               0.04 &  0.00 &    0.01 &       0.02 &      0.04 &     0.06\\
$\Delta($Eu/Fe$])$   & 0.01 &               0.03 &  0.02 &    0.04 &       0.03 &      0.06 &     0.07\\
$\Delta([$Dy/Fe$])$  & 0.04 &               0.05 &  0.03 &    0.02 &       0.01 &      0.07 &     0.09\\
\hline 
\end{tabular}
\label{tab:error}
\end{table*}

\section{Results}   
\label{sec:results}

Our work involves the most extensive measurement of abundances carried out in NGC 6809 in terms of number of elements considered. We aim to study the chemical evolution of NGC 6809 and to also look for possible MPs. In the following sections, we present chemical abundances for $\alpha$, Fe-peak, light and heavy elements. We compare our results with Halo and Disc field stars as well Galactic Globular Clusters that have NGC 6809's metallicity.

\subsection{Alpha Elements}

A common feature in almost all the Globular Cluster and Halo field stars as well as the metal-poor stars in the Milky Way, is an over-abundance of the $\alpha$ elements content. To date the only exception found among GCs is Rup. 106 \citep{Villanova_Rup106} which shows solar $\alpha$ elements abundances in spite having [Fe/H]=-1.5 dex.

\begin{figure}
  \includegraphics[width=1.01\columnwidth]{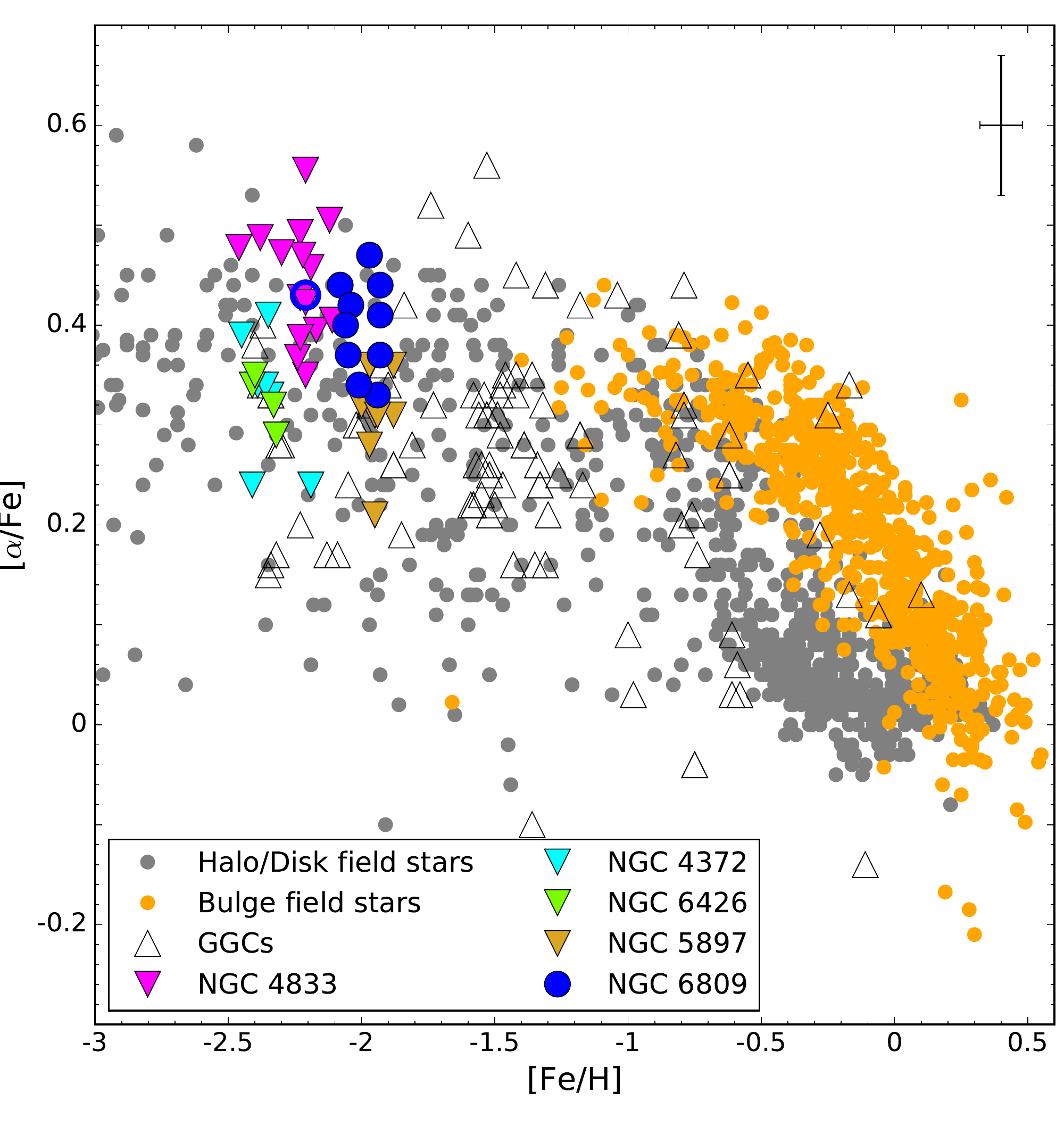}
\caption{[$\alpha$/Fe] ratio versus [Fe/H]. Filled blue circles are our data for NGC 6809. Open blue circle is our data for star \#1 (the most metal poor of our sample). Triangles represent different GCs samples: Empty black triangles: Galactic Globular cluster \citep{Pritzl_2005}. Filled green triangles: NGC 6426 \citep{Hanke_2017}. Filled cyan triangles: NGC 4372 (Valenzuela-Calderon et al. in prep). Filled magenta triangles: NGC 4833 \citep{Roederer_2015}.Filled golden triangles: NGC 5897 \citep{Koch_2104}. Filled orange circles: Bulge field stars \citep{Gonzalez_2011}. Filled gray circles: Halo and field stars \citep{Venn_2004,Cayrel_2004}.}
\label{fig:alpha} 
\end{figure}

\begin{figure}
    \includegraphics[width=1.07\columnwidth]{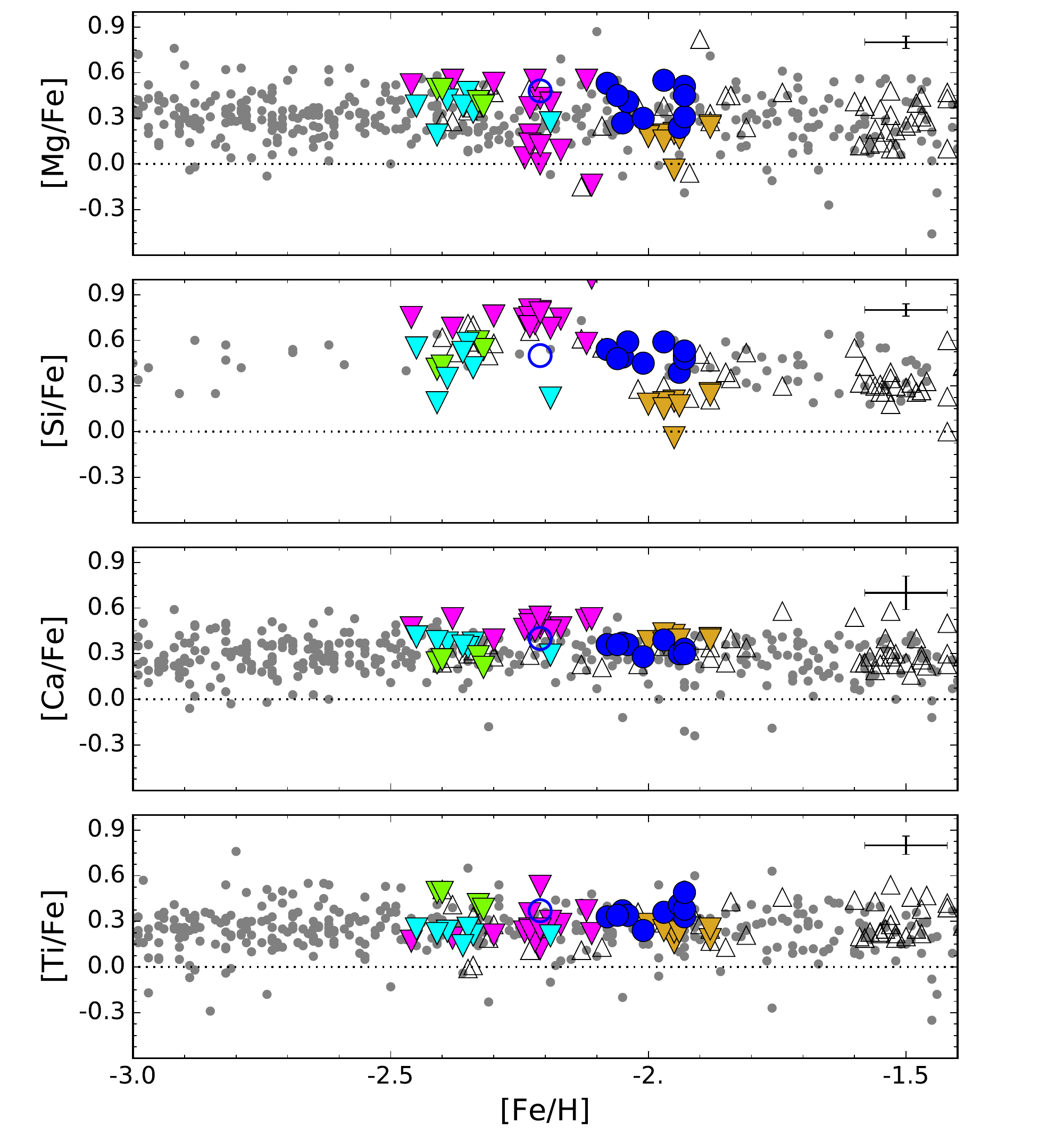}
    \caption{[Mg/Fe], [Si/Fe], [Ca/Fe] and [Ti/Fe] ratios versus [Fe/H]. Filled blue circles are our data for NGC 6809. Open blue circle is our data for star \#1 (the most metal poor of our sample). Triangles represent different GCs samples: Empty black triangles: Galactic Globular cluster \citep{Pritzl_2005}. Filled green triangles: NGC 6426 \citep{Hanke_2017}. Filled cyan triangles: NGC 4372 (Valenzuela-Calderon et al. in prep). Filled magenta triangles: NGC 4833 \citep{Roederer_2015}.Filled golden triangles: NGC 5897 \citep{Koch_2104}. Filled orange circles: Bulge field stars \citep{Gonzalez_2011,Johnson_2014}. Filled gray circles: Halo and field stars \citep{Venn_2004,Cayrel_2004,Barklem_2005}.}
    \label{fig:alpha_2}
\end{figure}

The $\alpha$ elements (O, Mg, Si, Ca, Ti) in our work are overabundant compared to the Sun. Ca and Ti are enhanced by $\sim$+0.30 dex, while Mg and Si by $\sim$+0.40 dex and $\sim$+0.50 dex respectively. Since O shows a star to star variation it will be treated separately in the Subsection \ref{Sub:Na-O}. Based only on Mg, Si, Ca and Ti we derived a mean $\alpha$ content of

$$\langle[\alpha/Fe]\rangle= +0.40 \pm 0.01$$ 

where the reported error is the error of the mean. Figure \ref{fig:alpha} shows the [$\alpha$/Fe] ratio as a function of the metallicity for each star in our sample, Galactic Globular Clusters (GGCs), Disc and Halo stars. In Figure \ref{fig:alpha_2} we plot the individual [Mg/Fe], [Si/Fe], [Ca/Fe] and [Ti/Fe] ratios as a function of the metallicity. Error bars are plotted in each panel. NGC 6809 follows the same trend as GGCs (e.g. NGC 5897, NGC 6426, NGC 4372 and NGC 4833) and is fully compatible with Halo field stars. The mean $\alpha$ content for each star is reported in Table \ref{tab:abundances}. 

As mentioned previously, the number of spectroscopic studies performed in NGC 6809 is really small. Comparing our results with previous studies performed in NGC 6809 we find only small differences in the pure $\alpha$-element Silicon. In particular [Si/Fe] abundances are $\sim$+0.12 dex higher than those found in \citet[hereafter C2009b]{CarrettaUves}\defcitealias{CarrettaUves}{C2009b}, but in agreement with the values found in \citet[hereafter PSG84]{Pilachowski_Alfa}\defcitealias{Pilachowski_Alfa}{PSG84}. In the case of Calcium our results are in good agreement with both \citetalias{Pilachowski_Alfa} and \citet[hereafter C2010]{Carretta_2010_Ca} \defcitealias{Carretta_2010_Ca}{C2010}. Additionally, for the pure $\alpha$ elements we found a mean value of [(Si+Ca)/Fe]= +0.41 this value is consistent with the mean value report above. For Titanium, our results for NGC 6809 are 0.14 dex higher than those found in \citetalias{Pilachowski_Alfa}. 

The explanation of the $\alpha$-enhancement observed in NGC 6809 is the classical SNeII pollution of the proto-cluster cloud stars. This type of supernovae is relatively an efficient producer of $\alpha$-elements \citep{Sneden_2004}.

Analyzing both the total observational error and the observed scatter for Si, Ca and Ti (see Table \ref{tab:error}), we conclude that there is no evidence of internal dispersion for these elements (dispersion in Mg will be treated separately in Subsection \ref{Sub:Mg-Al}).

\subsection{Light elements}
\label{Sub:Light_elements}
Strong evidence for the existence of Multiple Populations (MPs) in GC is the variations in the abundances of light elements Na, O, Mg and Al. These elements are involved in the CNO, NeNa, MgAl cycles of p-capture reactions related to H-burning at high-temperature showing a large star-to-star variation in GC clusters. Field stars with similar metallicities do not share this behaviour. On the other side, GCs contain stars that are characterised by the same abundance pattern observed in field stars of the same metallicity. This indicates that GCs are made up of MPs, as we already mentioned in Section \ref{Sec:intro}. In NGC 6809 these variations are present and in the following section, we discuss these relations in order to understand the MPs in this cluster.

\subsubsection{Na-O anticorrelation}
\label{Sub:Na-O}

Sodium and Oxygen show the well-known anticorrelation found in almost all GCs. This feature has been used to define a GC \citep{CarrettaGiraffe, Gratton_2012} and is caused by both proton-capture processes in H-burning at high temperatures through the CNO cycle which depleted Oxygen and the NeNa cycle which enriches Sodium \citep{Arnould_1999}. We found a pronounced spread in both Na and O abundances (see Tab. \ref{tab:error}) revealing the presence of this classical anticorrelation in our cluster.

\begin{figure*}
  \includegraphics[width=1.03\columnwidth]{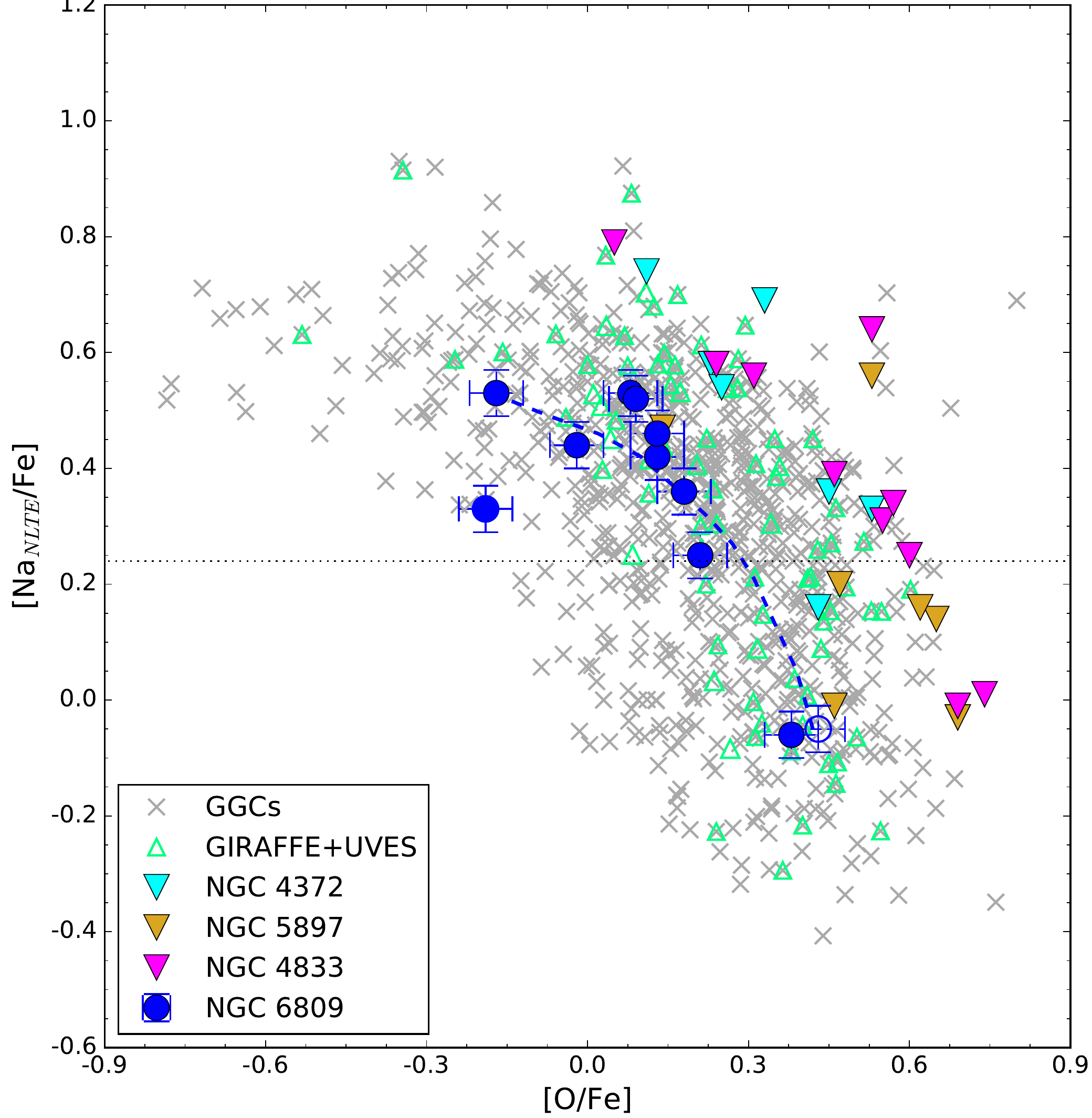}
   \includegraphics[width=1.03\columnwidth]{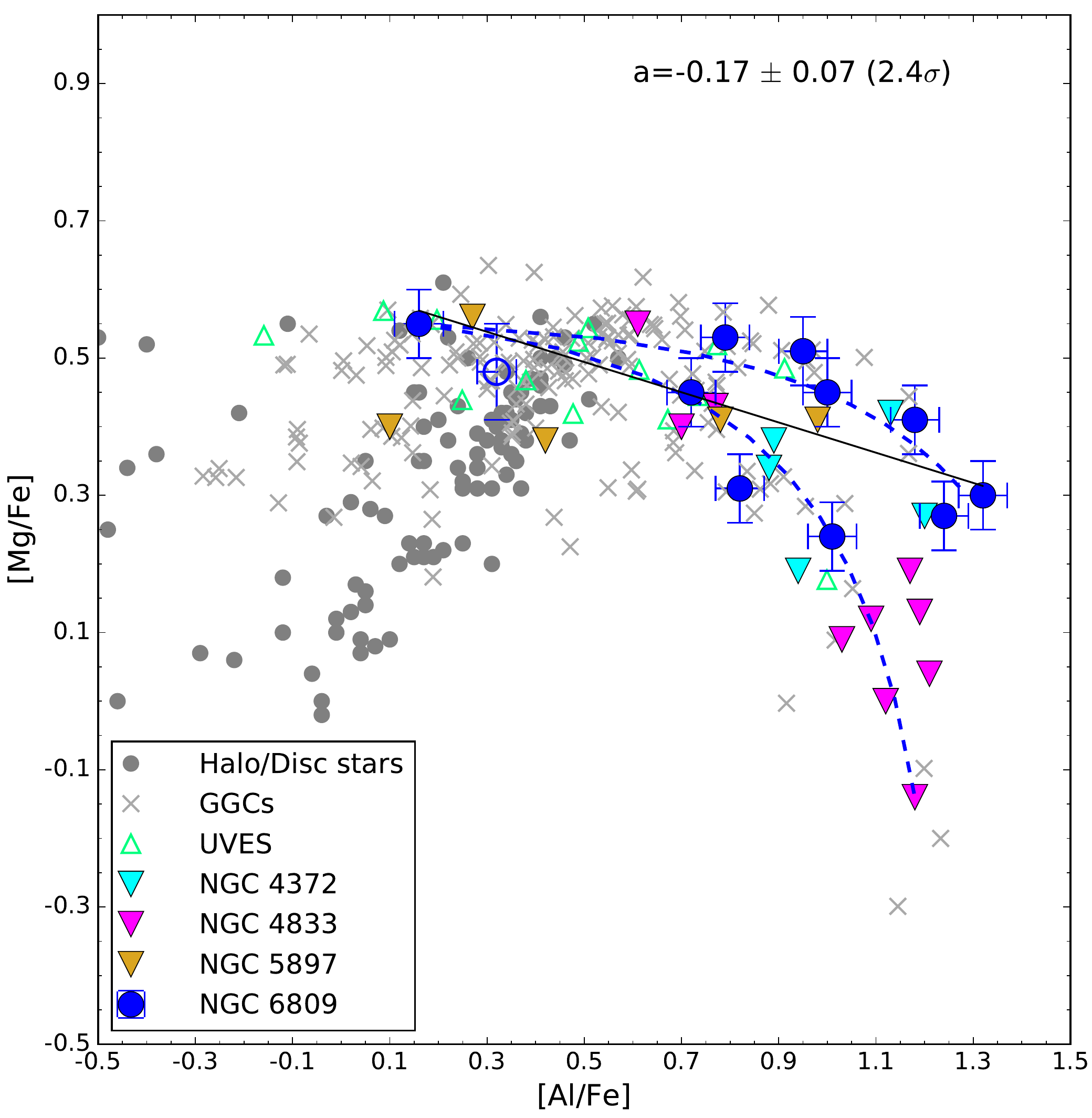}
\caption{\textbf{Right:} [Na/Fe] versus [O/Fe]. Open blue circle is our data for star \#1 (the most metal poor of our sample). Triangles represent different GCs samples: Open pale green triangles represent UVES and GIRAFFE data from \citet{CarrettaUves} for NGC 6809. Filled cyan triangles: NGC 4372 (Valenzuela-Calderon et al. in prep). Filled magenta triangles: NGC 4833 \citep{Roederer_2015}. Filled golden triangles: NGC 5897 \citep{Koch_2104}. Gray crosses: Galactic Globular Cluster \citep{CarrettaUves}.\textbf{Left:} [Mg/Fe] versus [Al/Fe]. Filled gray circles: Halo and Disc field stars \citep{Fulbright_2000, Cayrel_2004}. Black dash line is the best fit for our sample. Other symbols are the same as in the plot on the right.}
    \label{fig:na-o-anticorrelation} 
\end{figure*}

The only previous study looking for the Na-O anticorrelation in NGC 6809 was performed by \defcitealias{CarrettaGiraffe}{C2009a}\citetalias{CarrettaGiraffe, CarrettaUves}. In Figure \ref{fig:na-o-anticorrelation} we plot the [Na/Fe] values as a function of [O/Fe] found in each star of our sample together with those found in the paper recently mentioned. We considered the agreement satisfactory for both Na and O. In comparison with \citetalias{CarrettaUves} our dispersions of Oxygen are higher ($\sigma_{Obs}$ = 0.18 dex versus $\sigma_{Obs}$ = 0.11 dex) and our mean value is slightly lower. NLTE corrections were made for Na lines based on the INSPEC \footnote{http://inspect.coolstars19.com/index.php?n=Main.HomePage} database \citep{Lind_2011}. The mean NLTE corrections obtain for [Na/Fe] is $\sim$-0.08 dex.

\citet[hereafter W2017]{Wang}\defcitealias{Wang}{W2017} found a mean [Na/Fe] = 0.23 $\pm$ 0.15 for the UVES sample and [Na/Fe] = 0.27 $\pm$ 0.16 for the GIRAFFE sample, the differences with our work beeing 0.10 and 0.06 dex respectively. This offset probably reflects the difference of $\sim$0.1 dex we found for Iron with respect to this paper.

If we look at Figure \ref{fig:na-o-anticorrelation} it is possible to identify two stars (\#1 and \#3) in the Na-poor/O-rich region, related to the primordial (P) stellar component (below the dashed horizontal line). On the other hand, a group of Na-rich/O-poor stars provide evidence of second-generation GC stars among the sample. 

We fitted the dilution model following \cite{CarrettaUves}. This model fits well most of the stars, only star \#10 is left out. This star does not follow the Na-O anti-correlation described by the rest of the stars in NGC 6809 (see Figure \ref{fig:na-o-anticorrelation}, left panel). The low estimate of O and Na content could be explain by the fact that this star has the lowest S/N of the entire sample (see Table \ref{tab:basic_parameters})


\subsubsection{Mg-Al anticorrelation}
\label{Sub:Mg-Al}
To be able to investigate the MPs in NGC 6809, we also need to investigate the anticorrelation between Magnesium and Aluminum. This is useful because the first and second generations lead not only to variations in the Na and O abundance but also possible variations of the Al and Mg abundances. The origin of this variation in NGC 6809 is just the result of the Mg-Al cycle, which converts Mg into Al at a temperature of $\sim$ 70 million K \citep{Charbonnel_2009}.

 \begin{figure}
\includegraphics[width=1.03\columnwidth]{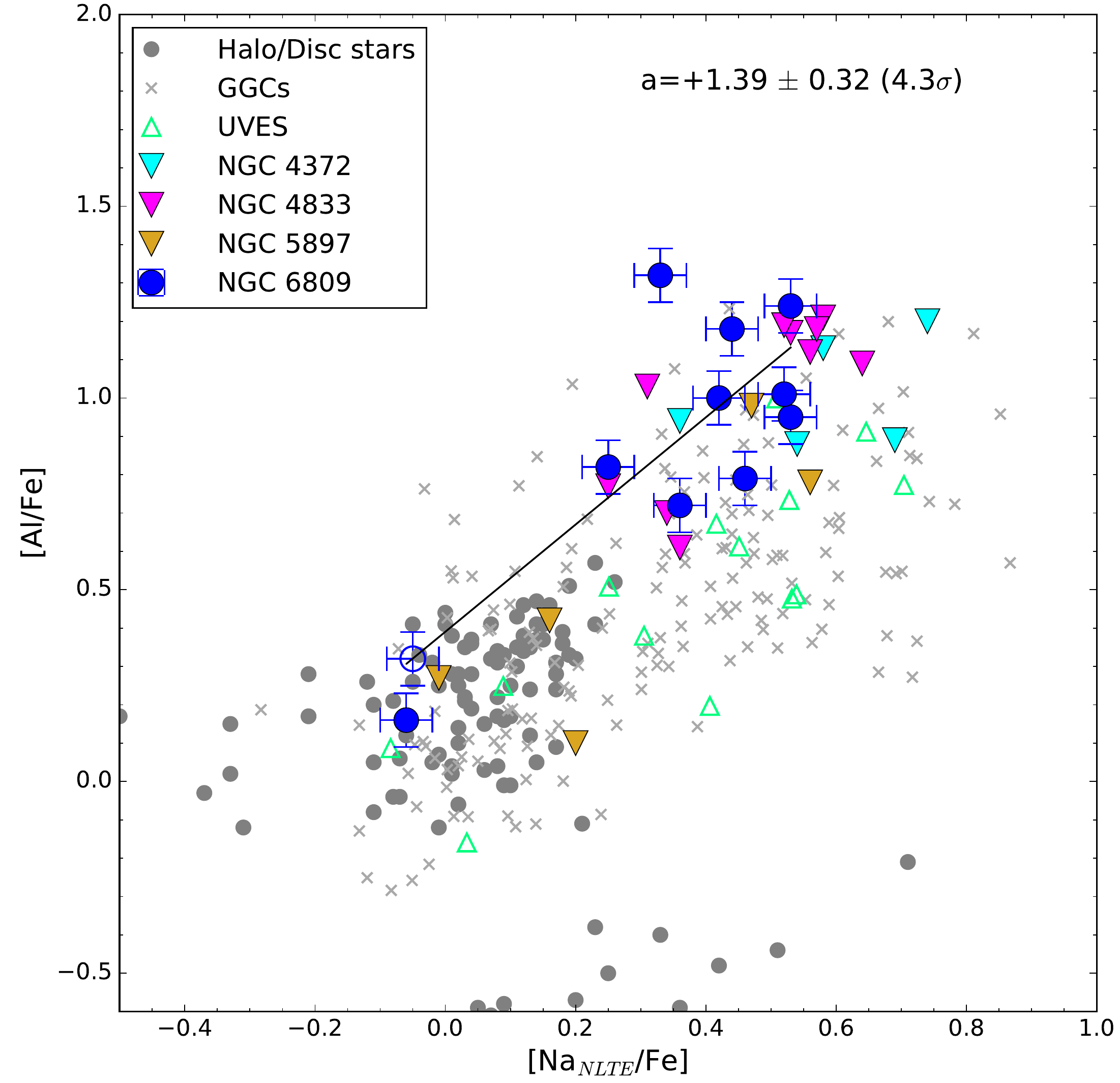}
\caption{[Al/Fe] versus [Na/Fe]. Filled blue circles are our data for NGC 6809. Open blue circle is our data for star \#1 (the most metal-poor of our sample). Triangles represent different GCs samples: Open pale green triangles represent UVES and GIRAFFE data from \citet{CarrettaUves} for NGC 6809. Filled cyan triangles: NGC 4372 (Valenzuela-Calderon et al. in prep). Filled magenta triangles: NGC 4833 \citep{Roederer_2015}. Filled golden triangles: NGC 5897 \citep{Koch_2104}. Gray crosses: Galactic Globular Cluster \citep{CarrettaUves}. Filled gray circles: Halo and Disc field stars \citep{Fulbright_2000, Cayrel_2004}. Black line is the best fit for our sample.}
    \label{fig:na-al} 
\end{figure} 

Both elements show an internal dispersion, being more clear for Aluminum ($\sigma_{Obs}$ = 0.34 v/s $\sigma_{Tot}$ = 0.07) than for Magnesium ($\sigma_{Obs}$ = 0.10 v/s $\sigma_{Tot}$ = 0.04). This is not surprising since in general Aluminum abundances show a high spread in almost all galactic GCs, especially in those with a metallicity lower than [Fe/H]=-1.1 dex \citep{CarrettaUves, Meszaros_2015, Pancino_2017}. We found that the differences between our most Al-rich and Al-poor \textbf{stars} is about $\sim$1.10 dex, on the other hand, the difference between our most Mg-rich and Mg-poor stars is $\sim$0.30 dex. The left panel of Figure \ref{fig:na-o-anticorrelation} displays the [Mg/Fe] ratio as a function of [Al/Fe] ratio together with data from \citetalias{CarrettaUves} and different GCs having similar metallicity to NGC 6809. 	


The anti-correlation found is clear. Additionally, we perform a linear fit to the sample; this fit has a slope of a = -0.17 with an error of $\pm$0.07. This corresponds to a significance of the slope of 2.4$\sigma$ that we consider statistically significant. However, it appears to us that a linear fit is not the best interpretation of the data. If we combine our results with those from other GCs, we see quite clearly that in Figure \ref{fig:na-o-anticorrelation} (left panel), two Mg-Al relation appears. These are indicated by the two dilution models we fit the data. In NGC 6809 both are present as in NGC 4372, while in other clusters like NGC 4833 only one of them is found. This could indicate that in NGC 6809 two kinds of polluters are responsible for the MP-phenomenon. More data are required to confirm this behaviour. With all the aforementioned evidence, we can confirm the existence of a Mg-Al anti-correlation in NGC 6809 for the first time.

\subsubsection{Na-Al correlation}
\label{sec:na-al}
Al is expected to correlate with elements enhanced by proton-capture reactions (e.g. Na). Several authors found a correlation between Aluminum and Sodium \citep[e.g][]{Shetrone_1996, Ivans_2001, CarrettaUves}. 
Features of these elements are in general easier to measure, and the associated changes in abundance are usually larger than for O and Mg. 

In Figure \ref{fig:na-al} we plot [Al/Fe] ratio as a function of [Na/Fe] ratio. 
The interesting point is that our correlation is not continuous, sharing the same behaviour shown in both Na-O and Mg-Al anticorrelations where our sample is categorised into two groups: two stars (\#1 and \#3) with low-Na and low-Al and nine stars with high-Na and high-Al. We perform the best-fit to our sample, this fit has a slope of a = +1.39 with an error of $\pm$0.32 this corresponds to a significance of the slope of 4.3$\sigma$. The correlation between Sodium and Aluminum is clear.

\subsection{Iron and iron-peak Elements}

The mean [Fe/H] value we found for this cluster is:
 
 $$\langle[Fe/H]\rangle=-2.01 \pm 0.02$$

where the reported error is the error of the mean. We did not find signs of intrinsic spread since the dispersion in iron values are within the errors (see Tab. \ref{tab:error}). We decided to mark star \#1 with a blue open circle in all the plots because this star has a difference of -0.2 dex compared to the average iron content.

Excluding star \#1 our mean iron content is [Fe/H]=-1.99$\pm$0.01. The most likely explanation for the behaviour of star \#1 could be that it probably is a pulsating variable. We found a similar case in NGC 6528, where one star significantly more iron poor than the rest of the sample was found \citep{Cesar_2018}. Using photometry from the VVV infrared survey, we could prove that this star is a pulsating variable. Despite having a lower iron content, all the abundance ratios were in good agreement with the remaining stars. Our star \#1 represents a similar case, although we lack the necessary photometry data in order to prove it.

Unlike $\alpha$-elements, it is easy to find spectroscopic measurements of metallicity for NGC 6809 stars. \cite{Minniti_1993} found a value of [Fe/H]=-1.95 dex by using only 2 stars, \cite{Carretta_metallicity} found from 14 stars observed with UVES a value of [Fe/H]=-1.93$\pm$0.07 and recently \citetalias{Wang} who selected the targets we analyze in this paper found a value of [Fe/H]=-1.86 dex using UVES and GIRAFFE data ($\sim$100 stars). \citetalias{Wang} measured iron abundances by using the same technique than us: equivalent width of both FeI and FeII unblended lines and the LTE program \texttt{MOOG}. The difference in iron content between \citetalias{Wang} and our study is about 0.1 dex. There are two main differences between our work and \citetalias{Wang} that probably introduced this off-set. They applied non-LTE corrections( 0.07 dex for the RGB sample and 0.10 dex for the AGB sample) to all the FeI lines and most of the sample in their work is made of GIRAFFE spectra that have lower resolution.

In our study we find that [V/Fe]=0.10 dex and  [Cr/Fe] and [Ni/Fe] ratios are solar scaled (0.0 dex), whereas the [Cu/Fe] and [Mn/Fe] ratios are underabundant both with a value of -0.28 dex.  [Sc/Fe], [Co/Fe] and [Zn/Fe] ratios are slightly overabundant with values of 0.13, 0.17 and 0.22 dex respectively. 
As in the case of low-metallicity GCs \citep{Simmerer_2003}, Copper is underabundant in NGC 6809. \cite{Koch_2104} give a value of [Cu/Fe]=-0.70 dex for NGC 5897 and \cite{Roederer_2015} found a value of [Cu/Fe]=-0.65 dex for NGC 4833. Cu abundance ratio in our study is moderately higher than for these GCs. In general, for Copper, stars follow a solar-scaled trend down to [Fe/H ]$\sim$ -0.9. Below that metallicity as in the case of NGC 6809 they drop to [Cu/Fe] $\sim$ -0.4. Cu is in agreement with the field anyway.

For -2.5 < [Fe/H] < -1.0 dex remains constant at -0.40 dex for field stars. In particular \cite{Sobeck_2006} found that for a sample of 19 GCs in the metallicity range -2.7<[Fe/H]<-0.7, the Mn abundances are similar to those of Halo field stars. They found a mean value of [Mn/Fe]=-0.37 dex for Globular Cluster stars and [Mn/Fe]=-0.36 for Halo field stars, our values for Mn values are in agreement with this finding and with those found for other GCs; e.g., our Mn content is in agreement with those found for NGC 4833 \citep{Roederer_2015} \citep{Carretta_2014} and NGC 6426 \citep{Hanke_2017}.

[Ni/Fe] is usually found to be close to the solar value over the entire metallicity range, suggesting that the origin of Ni is strictly linked to that of Fe from both SN types. Vanadium closely follow solar values, \cite{Gratton_1991} found [V/Fe] $\sim$ 0.0. Very similar to that of [Mn/Fe] is the behaviour of [Cr/Fe], which is found to increase with the metallicity.  Our cluster with [Ni/Fe]=0.0, agrees with the field star trend. The same occurs for [Cr/Fe] and [V/Fe], which closely follow field abundances.

\begin{figure}
    \includegraphics[width=1.07\columnwidth]{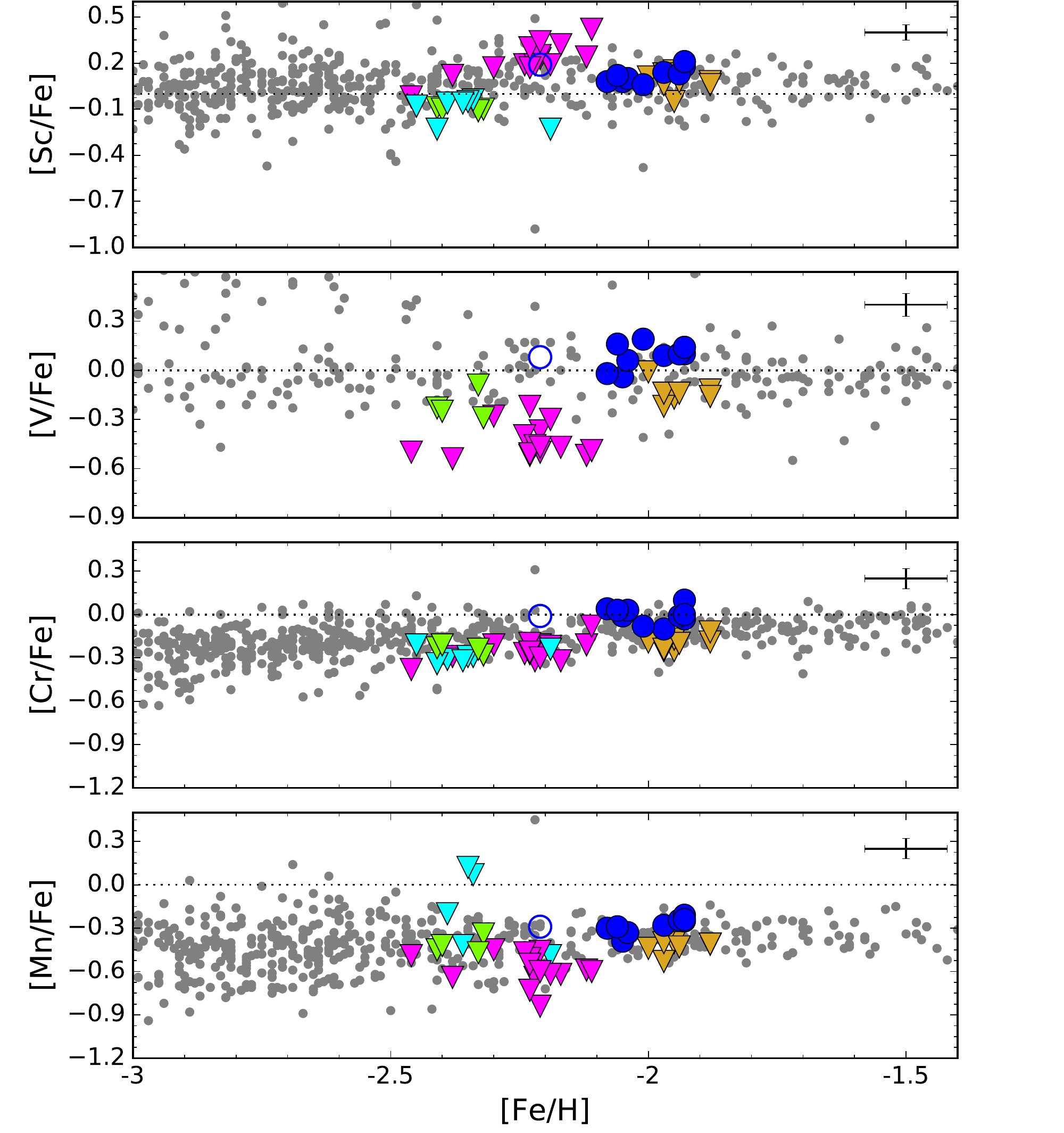}
    \includegraphics[width=1.07\columnwidth]{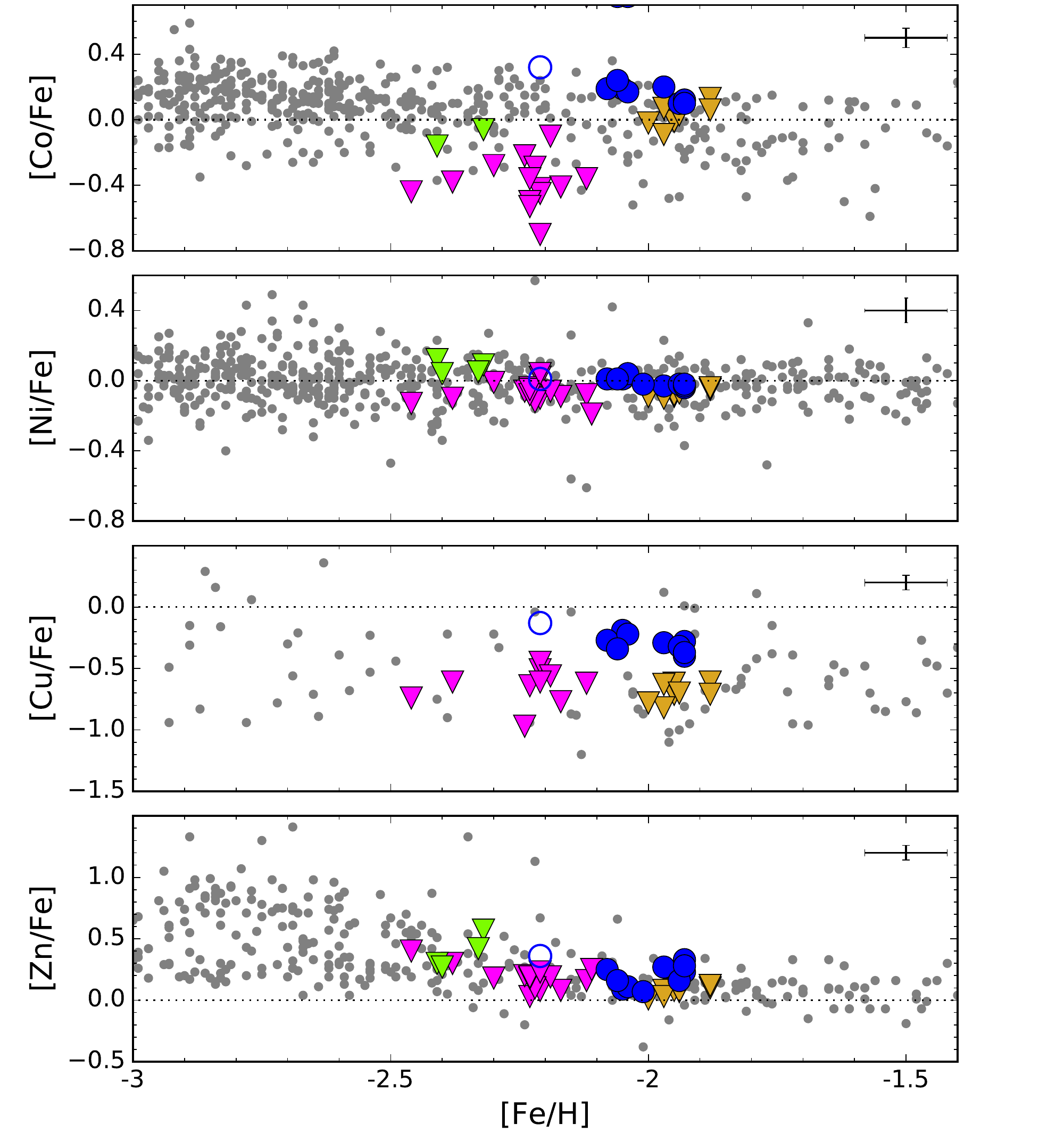}
    \caption{ [Sc/Fe], [V/Fe], [Cr/Fe], [Mn/Fe], [Co/Fe], [Ni/Fe], [Cu/Fe] and [Zn/Fe] ratios versus [Fe/H]. Filled blue circles are our data for NGC 6809.Open blue circle is our data for star \#1 (the most metal poor of our sample). Triangles represent different GCs samples: Filled green triangles: NGC 6426 \citep{Hanke_2017}. Filled cyan triangles: NGC 4372 (Valenzuela-Calderon et al. in prep). Filled magenta triangles: NGC 4833 \citep{Roederer_2015}. Filled golden triangles: NGC 5897 \citep{Koch_2104}. Filled gray circles: Halo and Disc field stars \citep{Fulbright_2000,Cayrel_2004,Barklem_2005,Ishigaki_2013, Roederer_2014}}
    \label{fig:iron_peak}
\end{figure}

Several author discussed if Scandium is solar scaled, that means [Sc/Fe]$\sim$0 at all metallicities \citep{Gratton_1991,MacWilliam_1995,Prochaska_2000} or if Sc behaves like an $\alpha$-element having [Sc/Fe]$\sim$+0.2 in metal-poor Halo and Thick-Disc stars \citep{Zhao_1990,Battistini_2015}. Our results lie in between (see Fig.\ref{fig:iron_peak}) so it is not possible for us to support any of these two possibilities. 

The chemical evolution of Co is thought to follow the same trend as Cr, being solar scaled or sub-solar at any metallicity. Observations show, however, that Co behaves like a $\alpha$-element in the sense that it is enhanced at low metallicities of about $\sim$+0.2 dex, decreasing towards higher metallicities \citep{Cayrel_2004, Ishigaki_2013, Battistini_2015}. The Co abundances in our sample are in agreement with the Co trend for Halo field stars. 

Observations of  Halo stars performed by \cite{Bensby_2005, Nissen_2007} show that in the range of -2.7 <[Fe/H]< -2 all [Zn/Fe] values are positive with an average value of [Zn/Fe] $\sim$ +0.15. 
According with \cite{Cayrel_2004} [Zn/Fe] rises steeply to $\sim$ +0.5 at the lowest metallicities. In our work we found a mean Zn content of 0.22 fully compatible with the galactic trend and GCs. 

Finally, analysing both the total observational error and the observed scatter for all the iron-peak elements (see Table \ref{tab:error}) we conclude that there is no evidence of internal dispersion for any of these elements.

\begin{figure*}
\includegraphics[width=2.\columnwidth]{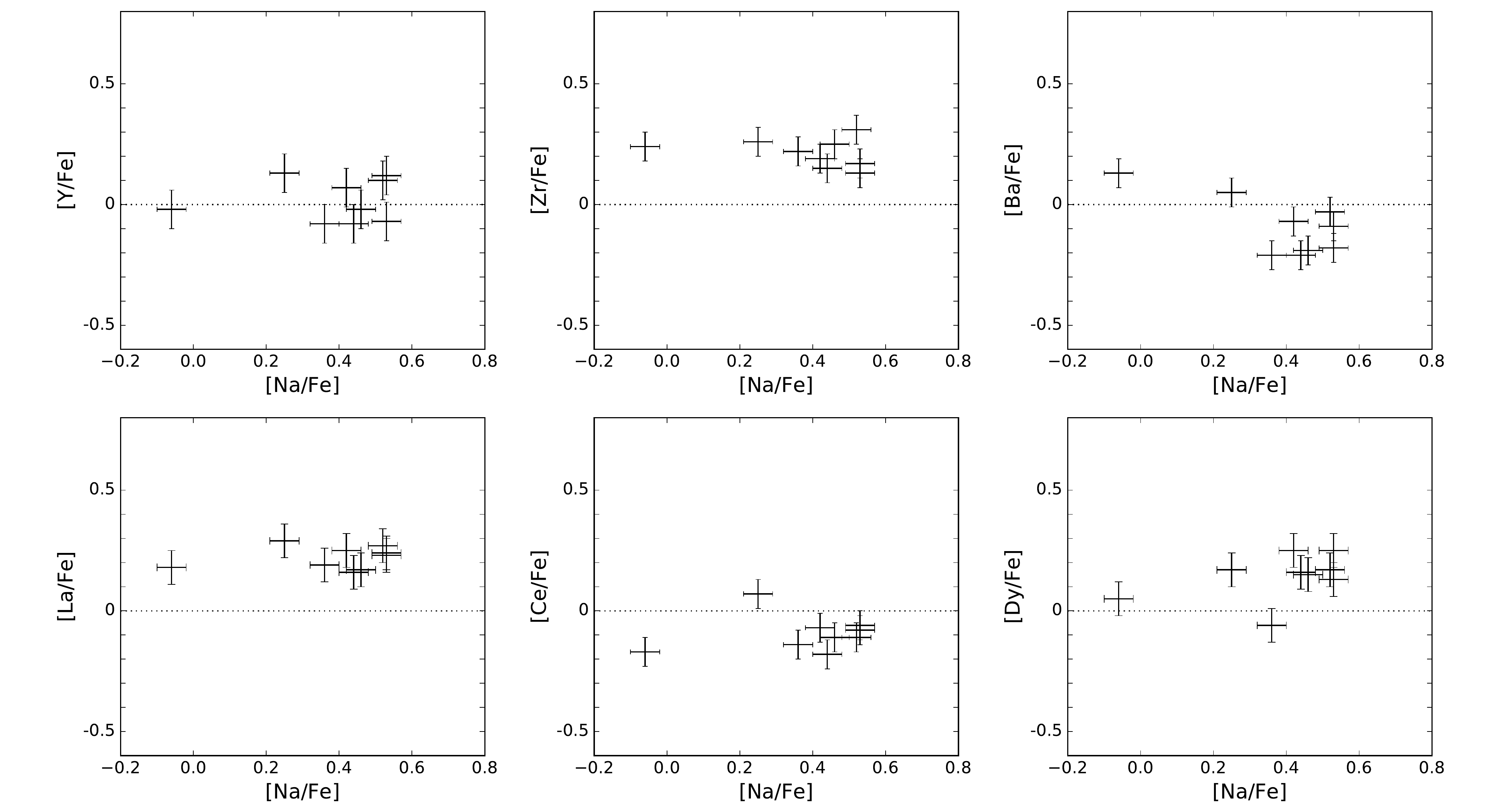}
\caption{ [Y/Fe], [Zr/Fe], [Ba/Fe], [La/Fe], [Ce/Fe] and [Dy/Fe] ratios versus [Na/Fe] ratio for our data.}
    \label{fig:s-elements} 
\end{figure*}

\subsection{Neutron capture Elements} 

We measured a number of elements with Z > 30: Y, Zr, Ba, La, Ce, Nd, Eu and Dy. These elements are mainly produced by neutron capture (n-capture) reactions through the s-process and the r-process. 
While the s-process is relatively well understood and occurs primarily in AGB stars, the r-process element formation is much more uncertain.
r-process element formation requires large neutron fluxes associated to rather 
catastrophic events. The two main candidates are type II (core-collapse) supernova explosions and neutron star mergers.The typical s-process elements observable in the stars are Sr, Y, Zr, Ba, and La, while the main chemical element produced through r-process is Eu.

\begin{figure}
\includegraphics[width=1.09\columnwidth]{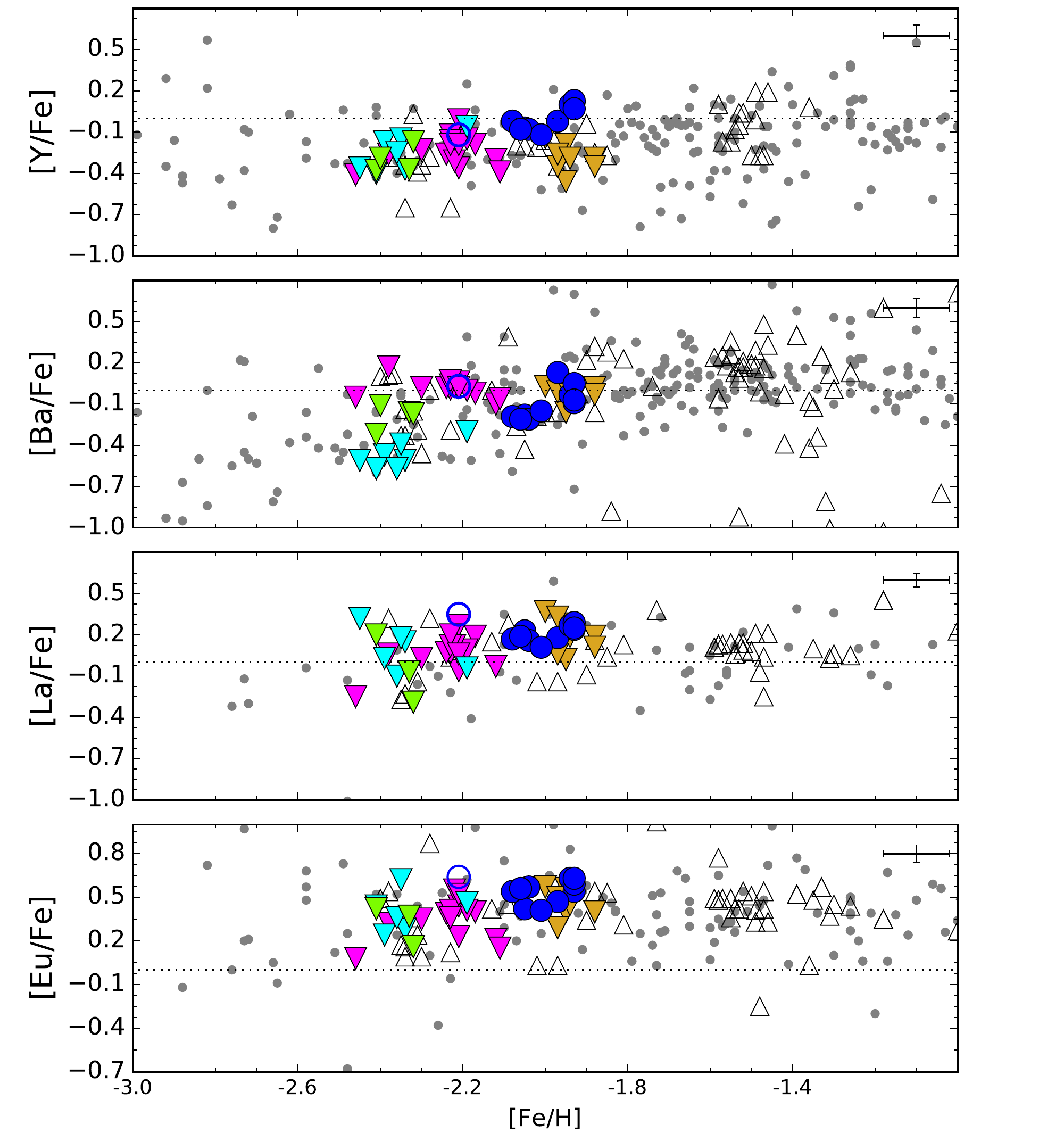}
\caption{[Y/Fe], [Ba/Fe], [La/Fe] and [Eu/Fe] ratios versus [Fe/H]. Filled blue circles are our data for NGC 6809. Open blue circle is our data for star \#1 (the most metal poor of our sample). Triangles represent different GCs samples: Filled green triangles: NGC 6426 \citep{Hanke_2017}. Filled cyan triangles: NGC 4372 (Valenzuela-Calderon et al. in prep). Filled magenta triangles: NGC 4833 \citep{Roederer_2015}. Filled golden triangles: NGC 5897 \citep{Koch_2104}. Filled gray circles: Halo and Disc field stars \citep{Fulbright_2000,Venn_2004}}
    \label{fig:heavy_all} 
\end{figure}

In Figure \ref{fig:heavy_all} we plot [Y/Fe], [Ba/Fe], [La/Fe] and [Eu/Fe] as a function of [Fe/H], together with GGCs around NGC 6809 metallicity, Halo and Disc field stars.

Both Yttrium and Barium shows high dispersions, $\sigma_{Obs}$ = 0.13 dex and $\sigma_{Obs}$ = 0.11 dex respectively, nevertheless both dispersions does not exceed significantly the errors and could be explained by the rms scatter. \cite{Ivans_2001} and \cite{Lee_2002} found that Barium shows a large dispersion at low metallicities. They also found a [Ba/Fe] ratios for giant stars in the range -0.30 < [Ba/Fe]  < +0.60 for GCs with -2.40 < [Fe/H] < -0.68. In this context our finding are fully compatible with these values \citep{James_2004}.

Europium is the most over-abundant n-capture elements ($\sim$0.50 dex) in our sample showing a smaller spread too ($\sigma_{Obs}$ = 0.07). Previous studies in GGCs give very similar results \citep[e.g][]{Ivans_2001, Ramirez_2003,Pritzl_2005,Muñoz_2013,Koch_2104,Hanke_2017}.

Lanthanum abundances in our sample have slightly over-abundant ($\sim$ +0.20 dex) values. Although literature values of La abundances are rare for metal-poor stars, it is well know that [La/Fe] ratio decreases with the metallicity for Halo field stars \citep{Ishigaki_2013}.

The neutron capture elements Dysprosium, Zirconium and Neodymium are slightly enhanced by +0.15, +0.22, and +0.35 dex respectively, while Cerium is under-abundant (-0.09 dex). None of these elements show significant dispersions. Their mean values are compatible with those of other GCs and field Halo stars of the same metallicity \citep[e.g.][]{Pritzl_2005, Koch_2104, Hanke_2017}. 

Heavy s-elements La and Ba show no correlation with any of the light elements that exhibites abundance variations; this is corroborated by the most Na-poor stars in our sample, that do not show clear signs of correlation. Additionally, we plot heavy and light s-elements  Ce, Y, Zr, La, Ba and Dy in Figure \ref{fig:s-elements} where none of these elements exhibits any significant correlations with the light element Na.

\begin{figure}
\includegraphics[width=1\columnwidth]{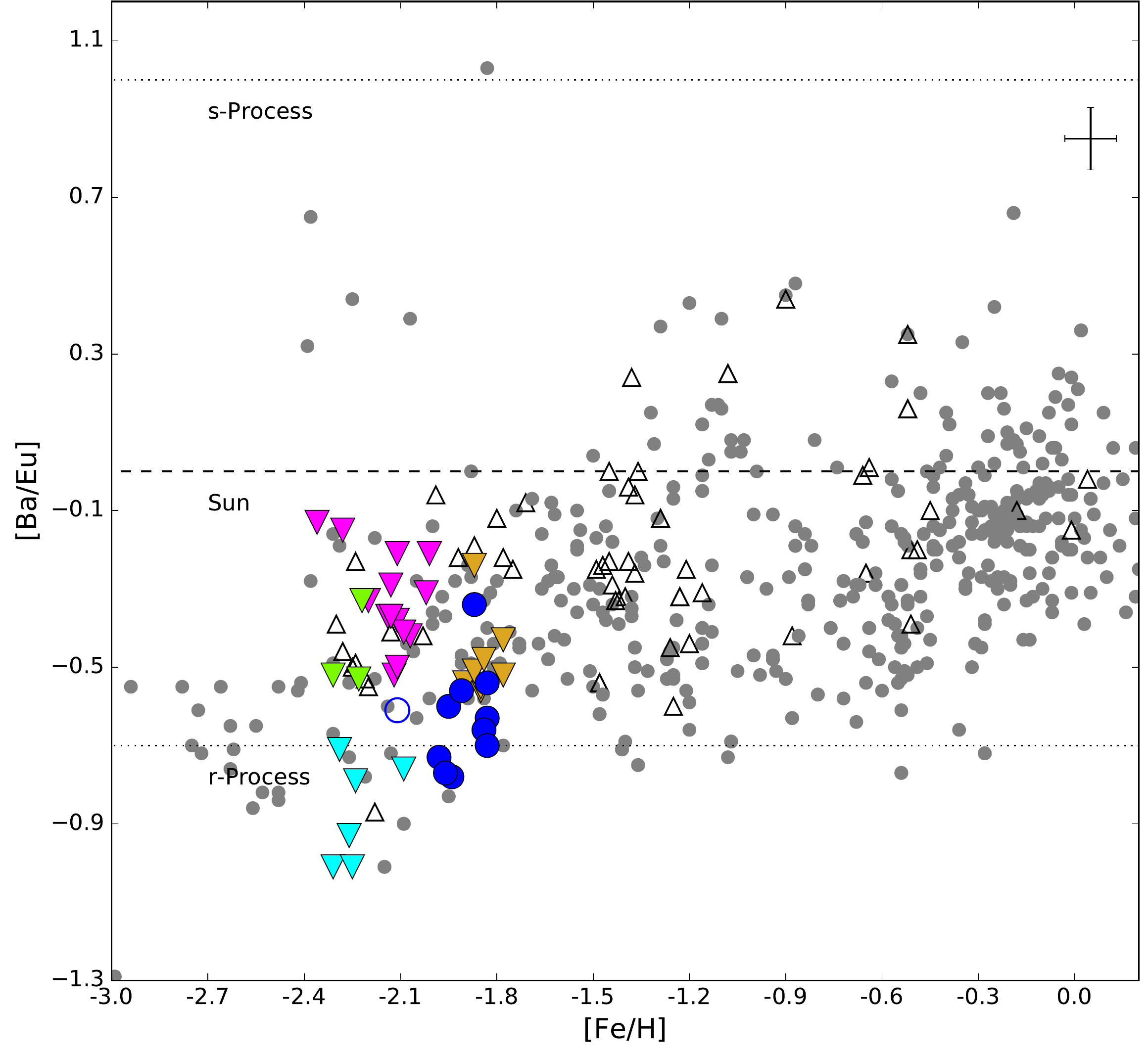}
\caption{[Ba/Eu] ratio versus [Fe/H]. Filled blue circle are our data for NGC 6809. Open blue circle is our data for star \#1 (the most metal poor of our sample). Triangles represent different GCs samples: Filled magenta triangles: NGC 4833 \citep{Roederer_2015}. Filled green triangles: NGC 6426 \citep{Hanke_2017}. Filled golden triangles: NGC 5897 \citep{Koch_2104}. Filled gray circles: Halo and Disc field stars \citep{Fulbright_2000,Venn_2004,Ishigaki_2013}}
    \label{fig:r_s_process} 
\end{figure}

The tracer for r-process nucleosynthesis is Eu, and those for the s-process are Ba and La. In order to know which process is involved in the enrichment of the proto-cluster cloud we plot in Figure \ref{fig:r_s_process} the [Ba/Eu] ratio as a function of [Fe/H]. We found a value of [Ba/Eu] which is very under-solar, and close to the pure r-process value [Ba/Eu]=-0.70 \citep{Arlandini_1999}. This result is fully compatible with previous analyses in this metallicity regime \citep{Fulbright_2000, Mashonkina_2003}. The previous result strongly suggests that heavy elements in NGC 6809 appear to have been produced by explosive events like core-collapse SNe or merging of neutron stars. The dispersion in [Ba/Eu] ratio is a direct consequence of the dispersion in Ba abundances (again, these dispersions are within the errors).

\section{Summary and Conclusions}
\label{sec:conclusions}

We have analysed UVES spectra of 11 stars of the Halo galactic Globular Cluster NGC 6809. The classical EW method was used for unblended lines, otherwise we applied the spectrum-synthesis method. We measured a mean cluster heliocentric radial velocity of  $\langle RV_{H}\rangle$ = 174.7 $\pm$ 3.26 kms$^{-1}$. For all the stars, we were able to perform a detailed chemical composition analysis. We found that:
\begin{description}
\item \textbf{$\alpha$-elements:} NGC 6809 shows the typical $\alpha$-enhanced pattern of the Halo with [$\alpha$/Fe] = +0.40 $\pm$ 0.04. Mg, Si, Ca and Ti are over-abundant.
According to our results, NGC 6809 suffered a rapid chemical evolution dominated by SNeII.\\

\item \textbf{Iron and iron-peak elements:} We found $\langle$[Fe/H]$\rangle$ = -2.01 $\pm$ 0.02 (error on the mean), in good agreement with previous studies \citetalias{Wang, CarrettaUves} and \cite{Harris}. No significant spread of iron is found in this work. In our study we find that [V/Fe], [Cr/Fe] and [Ni/Fe] ratios are solar scaled, whereas the [Cu/Fe] and [Mn/Fe] ratios are underabundant.  [Sc/Fe], [Co/Fe] and [Zn/Fe] ratios are slightly overabundant. The Fe-peak elements show good agreements with other Globular Clusters and Halo field stars. There is no evidence of internal dispersion for any of these elements.\\

\item \textbf{Heavy and Neutron capture elements:} The [Ba/Eu] ratio confirms a dominant contribution of the r-process ratios. [Y/Fe] is solar scaled. [La/Fe], [Zr/Fe], [Nd/Fe], [Eu/Fe] and [Dy/Fe] are overabundant. [Ce/Fe] and [Ba/Fe] are underabundant.\\

\item\textbf{Light elements:} We found that NGC 6809 exhibits the classical light element abundance variations associated with globular clusters, including the expected O-Na anti-correlation and for the first time we confirmed the presence of a Mg-Al anti-correlation that appears to be split into two sequences. We also found a Na-Al correlation. Al spread is the highest among the light elements. In our analysis, it is possible to observe two groups of stars: Two Na-poor/O-rich and Nine Na-rich/O-poor. The same behaviour is observed in the Mg-Al and Na-Al (anti)correlations.  
\end{description}

\section*{Acknowledgements}

MJ.R gratefully acknowledges the support provided by Direcci\'on de Postgrado UdeC and Chilean BASAL Centro de Excelencia en Astrof\'isica y Tecnolo\'ias Afines (CATA) grant PFB-06/2007 for international conference which helped to develop the paper.
SV and MJ.R gratefully acknowledges the support provided by Fondecyt reg. n. 1170518. 



\bibliographystyle{mnras}
\bibliography{bib}




\appendix

\bsp    
\label{lastpage}
\end{document}